\documentclass[useAMS,usenatbib]{mn2e}
\usepackage{graphicx} \usepackage{rotating}
\usepackage{txfonts}
\usepackage{url} 
\usepackage{fmtcount}
\usepackage{upgreek}

\def\aap{A\&A}

\def\aj{AJ}

\def\apj{ApJ}

\def\nt/f{Nuclear Technology/Fusion}

\title[PAH Formation in O-rich PNe]
{\begin{center}
PAH Formation in O-rich Planetary Nebulae
\end{center}}
\author[L. Guzman-Ramirez, E. Lagadec, D. Jones, A.~A. Zijlstra and K. Gesicki]{L. Guzman-Ramirez$^{1}$\thanks{E-mail: guzmanl@eso.org}\thanks{Based  on observations collected at the European Organisation for Astronomical Research in the Southern Hemisphere, Chile (PID: 087.D-0270)}, E. Lagadec$^{3}$, D. Jones$^{1,4}$, A.~A. Zijlstra$^{2}$ and K. Gesicki$^{5}$\\
$^{1}$European Southern Observatory, Alonso de C\'ordova 3107, Casilla 19001, Santiago, Chile\\
$^{2}$Jodrell Bank Centre for Astrophysics, School of Physics and Astronomy, University of Manchester, Manchester, M13 9PL, UK\\
$^{3}$Astronomy department, Cornell University, Ithaca, NY 14853-6801, USA\\
$^{4}$Universidad de Atacama, Copayapu 485, Copiap\'o, Chile\\
$^{5}$Centrum Astronomii UMK, ul.Gagarina 11, 87-100  Torun, Poland}

\date{Released 2014 Xxxxx XX}

\pagerange{\pageref{firstpage}--\pageref{lastpage}} \pubyear{2014}

\def\LaTeX{L\kern-.36em\raise.3ex\hbox{a}\kern-.15em
    T\kern-.1667em\lower.7ex\hbox{E}\kern-.125emX}

\begin{document}

\date{Accepted 2014 March 6.  Received 2014 March 6; in original form 2013 March 11}

\pagerange{\pageref{firstpage}--\pageref{lastpage}} \pubyear{2014}

\maketitle

\label{firstpage}

\begin{abstract}

Polycyclic aromatic hydrocarbons (PAHs) have been observed in O-rich planetary nebulae towards the Galactic Bulge. This combination of oxygen-rich and carbon-rich material, 
known as dual-dust or mixed chemistry, is not expected to be seen around such objects. We recently proposed that PAHs could be formed from the photodissociation
of CO in dense tori. In this work, using VISIR/{\it VLT}, we spatially resolved the emission of the PAH bands and ionised emission from the [SIV] line, confirming the presence of 
 dense central tori in all the observed O-rich objects. Furthermore, we show that  for most of the objects, PAHs are located at the outer edge of these dense/compact tori, while the ionised
 material is mostly present in the inner parts of these tori, consistent with our hypothesis for the formation of PAHs in these systems. The presence of a dense
 torus has been strongly associated with the action of a central binary star and, as such, the rich chemistry seen in these regions may also be  related to the
 formation of exoplanets in post-common-envelope binary systems.

\end{abstract}

\begin{keywords}
circumstellar matter -- infrared: stars, planetary nebulae.
\end{keywords}

\section{Introduction}

All low and intermediate-mass stars, with initial masses between 0.8 and 8 M$_{\odot}$, end their life as a white dwarf (WD). Some of them will 
experience the planetary nebulae (PNe) phase before the end of their lives, when the ultraviolet emission from
 the hot WD photodissociates and photoionises the expanding gas and dust that was ejected in the previous phase, when the star is on the asymptotic
 giant branch (AGB). During the AGB phase the intense mass loss (from $10^{-6}$ to $10^{-4}\,\rm M_{\odot}\, yr^{-1}$) leads to the formation of a circumstellar
 envelope made of gas and dust \citep{kwok}. The dust is seen in emission as an infrared excess superposed on the stellar spectral energy distribution (SED). 

In the AGB phase, a star may evolve from being oxygen-rich to being carbon-rich. The change occurs when carbon produced by He-burning is brought to 
the surface by dredge-up processes in the stellar interior thereby increasing the C/O ratio until it exceeds unity and a carbon star is formed \citep{herwig}.
  This process depends on stellar mass: \cite{vassi} showed the third dredge-up occurs if the core mass of the star $M_{cs} >
1.5$\,M$_{\odot}$. This predicts a clear distinction, with some (lower mass) stars showing oxygen-rich and some (higher mass) stars carbon-rich ejecta. 

In the molecular ejecta, the CO molecule locks away the less abundant element, leaving the remaining free O or C to drive the chemistry and dust formation.
  Oxygen-rich shells are characterised by silicate dust. Amorphous silicates dominate the 7-25$\upmu$m region in the IR spectrum, the 9.7$\upmu$m feature 
and the 18$\upmu$m are very strong bands that occur in emission  or (self-)absorption. Crystalline silicate emission features occur near 28, 33 and 43$\upmu$m \citep{sylvester}.
 Carbon-rich shells show Polycyclic Aromatic Hydrocarbon (PAH) emission bands and carbonaceous amorphous dust. The PAH emission bands are at 6.2, 7.7, 8.6, and 11.3$\upmu$m  \citep{leger84}.

\cite{zijlstra91} observed the first evidence for mixed chemistry when one planetary nebula (PN)  (IRAS07027-7934) with strong PAH emission bands, was found to also show a 1.6GHz OH maser line. Observations made with the Infrared Space Observatory ({\it ISO}) uncovered several further cases, where PAH emission in PNe occurred together with emission bands of silicates usually found in O-rich shells (Waters et al. 1998a,b; Cohen et al., 1999,2002)\nocite{cohen99,cohen02}. These PNe were shown to be the envelope of  late/cool [WC] type stars and the mixed chemistry was therefore explained as an evolutionary change in the central star - a recent thermal pulse, which made the star turn from O-rich to C-rich. This is the so-called very late thermal pulse (VLTP). In all these objects the torus remained O-rich, while the C-rich material is observed in the outflows. 

\cite{guten} and \cite{perea09} showed that the mixed chemistry phenomenon is widespread amongst Galactic bulge (GB) PNe. Their {\it Spitzer} observations show that the simultaneous presence of O and C-rich dust features is common, and is not restricted to objects with late/cool [WC] type stars. The traditional explanation relating mixed chemistry to a recent evolution towards carbon-star is highly unlikely for Bulge objects, as these old, low-mass stars are not expected to show substantial third dredge-up, and therefore should not show enhanced C/O ratios. The few AGB carbon stars in the Bulge do not originate from third dredge-up \citep{azzo}.  \cite{stephan} found that 4 AGB stars out of a sample of 27 objects in the GB show Tc, evidence of a third dredge-up. However, they also observed 45 AGB stars in the GB finding they are all O-rich \citep{stephan2}. The third dredge-up is thus occurring in some of these objects, but is not large enough to form carbon-rich AGB stars. 

\cite{me11} analysed a sample of 40 GBPNe, with the mixed chemistry phenomenon found in 30 nebulae. {\it HST} images and UVES spectra showed that the mixed chemistry is not related to the presence of emission-line stars, as it is in the Galactic disk population. Instead, a strong correlation is found with morphology, and the presence of a dense dusty equatorial structure (torus). The mixed chemistry phenomenon occurring in the GBPNe is best explained through hydrocarbon chemistry in an UV-irradiated, dense torus.  One way to test this theory is to spatially resolve the PAHs in these PNe. If the PAHs are only present in the outflows, this would support the hypothesis that they originate from a VLTP, implying that the central star changed from O-rich to C-rich. On the other hand, if the PAHs are concentrated in the torus, this would be more consistent with their formation resulting from the photodissociation of CO, meaning that the central star does not have to experience a third dredge-up nor VLTP to be able to produce C-rich molecules.  

Aiming to detect the PAHs either in the torus or the outflows, we selected 11 targets to be observed using the {\it VLT} spectrometer and imager for the mid--infrared (VISIR) instrument on the Very Large Telescope ({\it VLT}). We obtained images in three filters, PAH1 (8.59$\upmu$m), PAH2 (11.25$\upmu$m) and SIV (10.49$\upmu$m). We also used the long-slit spectrograph to analyse in more detail a sub-sample of three objects.

This paper is organised as follows. In Section 2, we present the observing technique. In Section 3, we show the results of the observations of 11 PNe, presenting the acquired images and spectra. In Section 4, we discuss the main results before presenting the final conclusions in Section 5.

\section{Observations}
The targets were selected based on their morphology in H$\alpha$; bipolar outflows with a torus structure, and an IRAS 10$\mu$m flux larger than 350mJy. Only 11 targets were selected due to time constraints on the instrument.
These 11 targets are a sub-sample of the targets presented in \cite{me11}. For this original sample, the selection was based only on their flux (an IRAS 10$\mu$m flux larger than 350mJy). This selection criterion leads to a bias towards bipolar nebulae with equatorial structures where the dust is stable and hot, and thus bright in the mid-infrared. This effect is also seen in the MIR observations of post-AGB objects from \citet{lagadec11}.

The observations were performed using the European Southern Observatory (ESO) mid-infrared instrument VISIR \citep{lagage04} installed on UT3 at the {\it VLT} (Paranal, Chile).
 This instrument is composed of an imager and a long-slit spectrometer covering several filters in  N (8-13$\upmu$m) and Q (16-24$\upmu$m) atmospheric windows.  Table \ref{log} presents the log of the observations.

\subsection{VISIR imaging}
The pixel scale of VISIR in the imaging mode employed is $0\farcs0755$ and the field of view $19\farcs2 \times 19\farcs2$. These observations were done using the standard VISIR imaging mode, with the standard chopping and nodding technique to remove the sky background. This chopping technique uses the moving secondary mirror of the telescope in order to beam switch between target and sky (a chop-throw of 10$^{\prime\prime}$ was used for these observations). All the frames taken at a given chopping position are co-added immediately at the end of the exposure. The detector integration time (DIT) for a single frame is typically of the order of 10/20ms. To reduce the data, we used the standard ESO pipeline as integrated into the Gasgano software (version 3.4.4). This pipeline applies flat field correction, bad pixel removal, source alignment and co-addition of the frames for each filter. Images were obtained with three filters: PAH1, PAH2, and SIV (8.59 $\pm$ 0.42$\upmu$m, 11.25 $\pm$ 0.59$\upmu$m, and 10.49 $\pm$ 0.16$\upmu$m, respectively). The PAH1 filter's transmission covers the 8.6$\upmu$m PAH, SIV includes the [SIV] emission line and the PAH2 filter covers the 11.3$\upmu$m PAH feature. All filters include a contribution from the continuum. The continuum contribution varies from 5--30\% depending on the object, this was measured using the N band of the VISIR spectra (see Section 3.2).
According to the ESO manual\footnote{www.eso.org/sci/facilities/paranal/instruments/visir/doc} systematic uncertainties of VISIR photometric data are of the order of 10\% for PAH1 and PAH2.

 \subsection{VISIR spectroscopy} 
 We also acquired VISIR low-resolution spectroscopy (R $\sim$ 350 at 10$\upmu$m) from three PNe; M1-31, M1-40, and M3-38  from 9.0 to 13.4$\upmu$m split in three spectral segments of the N band, centred at 9.8, 11.4 and 12.4$\upmu$m. We used standard stars observed shortly before or after the observations to correct our spectra for telluric absorption. Extraction of the spectra, wavelength calibration, telluric line correction, and flux calibration were then carried out using a combination of the ESO provided pipeline and self-developed IDL routines. The long-slit spectrograph gave us a spatially resolved 2D spectrum for each object, from which we extracted the PAH features, the [SIV] emission line and the continuum. The spatial resolution of the VISIR spectrograph lies between 0.5-0.7\arcsec, depending on the seeing.

\begin{table*}
\caption{Log of our VISIR/{\it VLT} observations.}
\label{log}
\centering
\begin{tabular}{cccccccc} 
\hline\hline 
Name & PNG Name &  R.A.  & DEC & IRAS 12$\upmu$m flux  & Imaging/Spectroscopy &  Filter & Integration time \\
& & & &Jy & & & s\\
 \hline
Cn1-5 & 002.2-09.4 &  18 29 11.67& $-$31 29 58.8 & 1.69 & Imaging & PAH1& 300  \\
      &            &            &                &      & `` & PAH2& 1200 \\
      &            &            &                &      & `` & SIV & 1200 \\
M1-25 & 004.9+04.9 &  17 38 30.32& $-$22 08 38.8 & 0.78 & Imaging & PAH1& 1200  \\
      &            &            &                &      & `` & PAH2& 1200 \\
      &            &            &                &      & `` & SIV & 1200 \\
M1-31 & 006.4+02.0 &  17 52 41.44& $-$22 21 57.0 & 1.17 & Imaging & PAH2& 240  \\
      &            &            &                &      & `` & SIV & 240  \\
      &            &            &                &      & Spectrum & N & 600 \\
H1-61 & 006.5-03.1 &  18 12 33.99& $-$24 50 00.5 & 1.02 & Imaging & PAH2& 240 \\
      &            &            &                &      & `` & SIV & 240 \\	
M3-15 & 006.8+04.1 &  17 45 31.74& $-$20 58 01.8 & 0.53 & Imaging & PAH1& 240  \\
      &            &            &                &      & `` & PAH2& 240  \\
      &            &            &                &      & `` & SIV & 120  \\
Hb6   & 007.2+01.8 &  17 55 07.02& $-$21 44 40.0 & 1.88 & Imaging & PAH2& 240 \\
      &            &            &                &      & `` & SIV & 240 \\
M1-40 & 008.3-01.1 &  18 08 25.99& $-$22 16 53.2 & 2.70 & Imaging & PAH2& 960 \\
      &            &            &                &      & `` & SIV & 480 \\
      &            &            &                &      & Spectrum & N & 600 \\
Th3-4 & 354.5+03.3 &  17 18 51.94& $-$31 39 06.5 & 1.30 & Imaging & PAH1& 120 \\
      &            &            &                &      & `` & PAH2& 120 \\
      &            &            &                &      & `` & SIV & 120 \\
M3-38 & 356.9+04.4 &  17 21 04.46& $-$29 02 59.2 & 1.20 & Imaging & PAH1& 120 \\
      &            &            &                &      & `` & PAH2& 120 \\
      &            &            &                &      & `` & SIV & 120 \\
      &            &            &                &      & Spectrum & N & 360 \\
H1-43 & 357.1-04.7 &  17 58 14.44& $-$33 47 37.5 & 0.49 & Imaging & PAH1& 240 \\
      &            &            &                &      & `` & PAH2& 240 \\
      &            &            &                &      & `` & SIV & 480 \\
H1-40 & 359.7-02.6 &  17 55 36.05& $-$30 33 32.3 & 2.38 & Imaging & PAH1& 120  \\
      &            &            &                &      & `` & PAH2& 60 \\
      &            &            &                &      & `` & SIV & 120 \\
 \hline
   \end{tabular}
\end{table*}

\section{Results}

\begin{figure*}
\centering
\includegraphics[width=13cm]{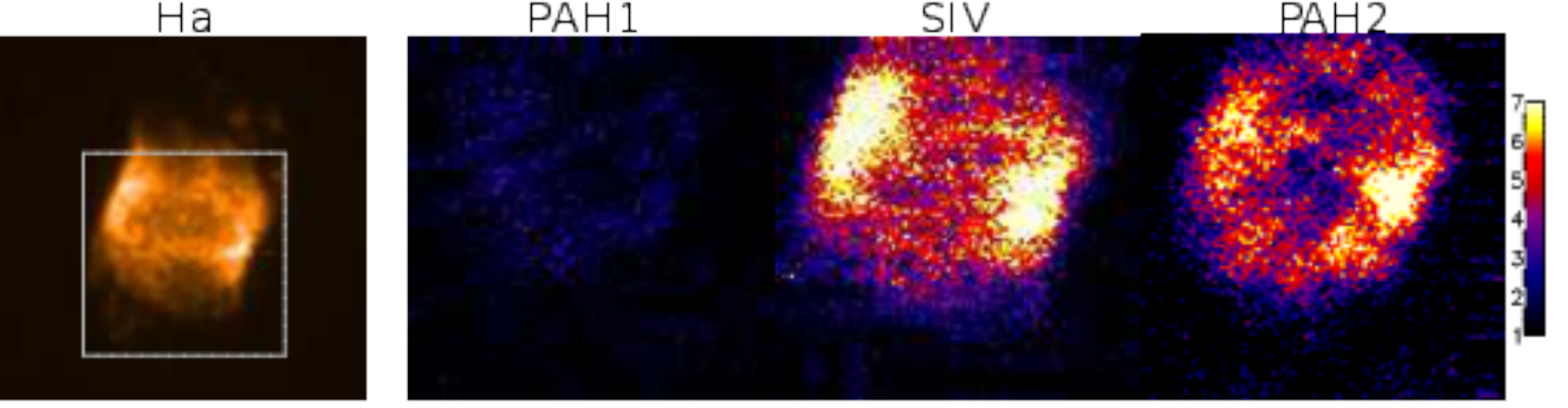}
\hbox{
\centering
\includegraphics[width=5.2cm]{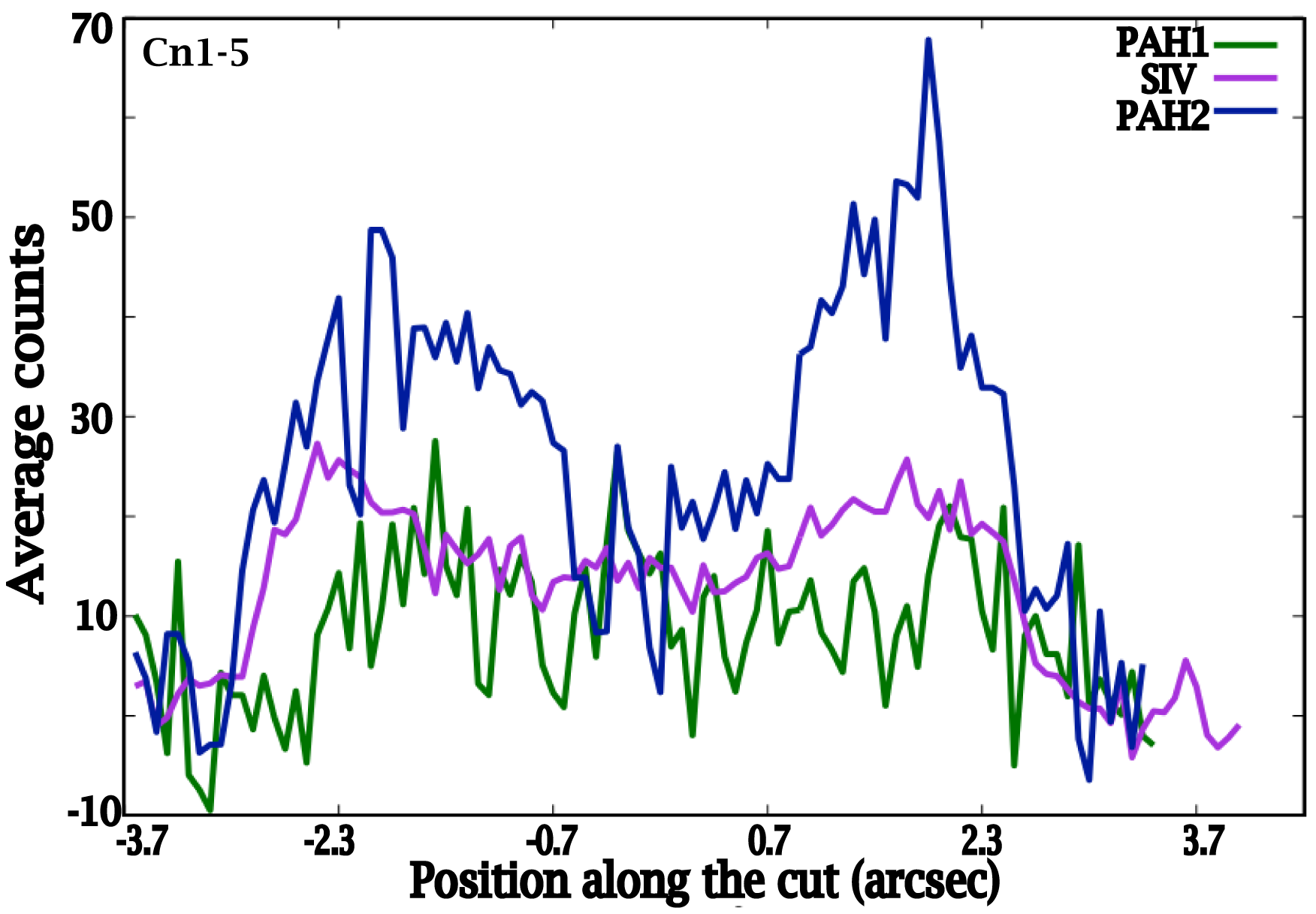}
\includegraphics[width=5.2cm]{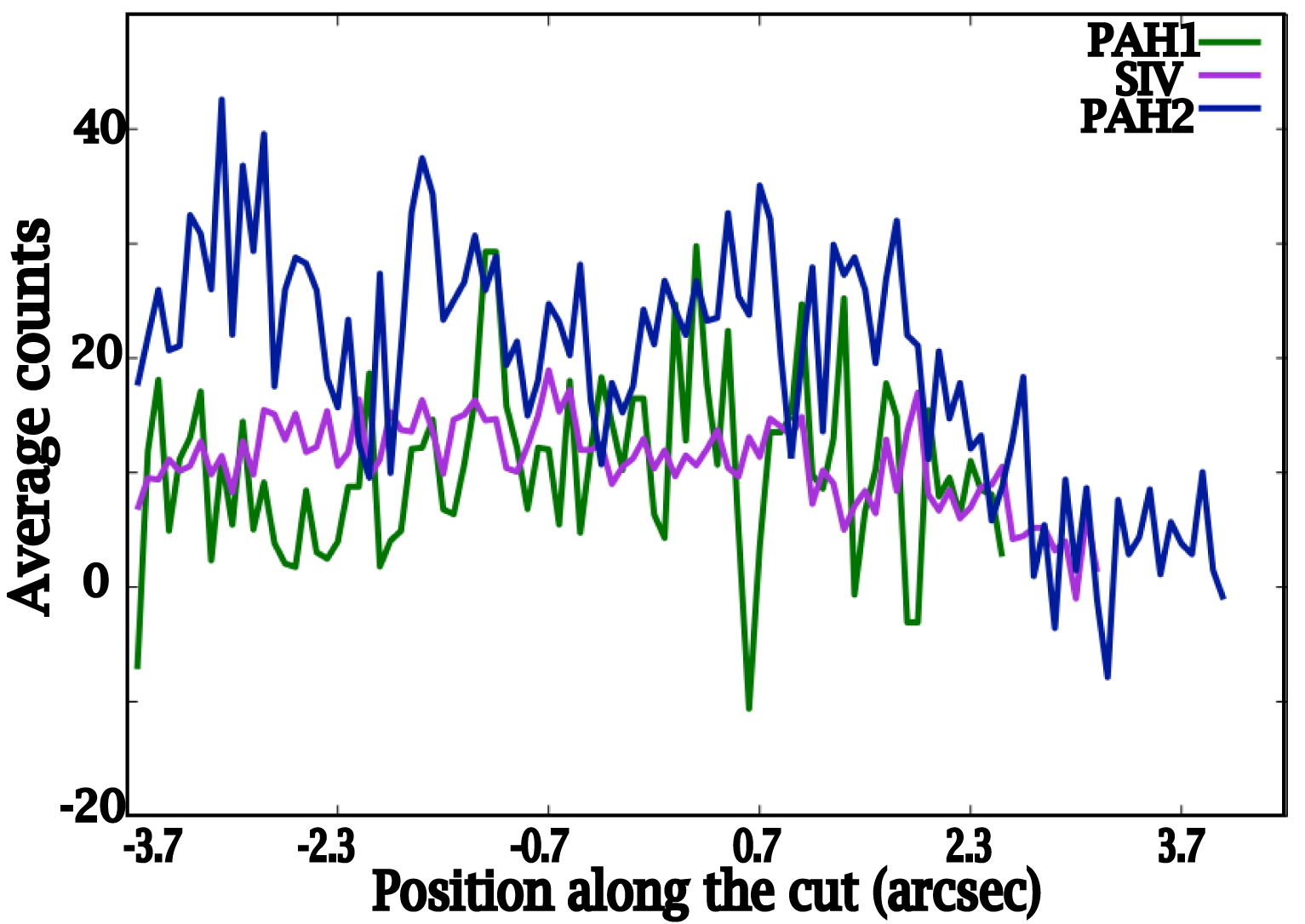}}
\caption[VISIR image of Cn1-5]{\label{cn1-5}PN Cn1-5. The leftmost image in the upper panel is an H$\alpha$ image taken with the {\it HST}, where the overlaid box shows the field observed by VISIR. The images to the right are (left to right): VISIR observations at 8.59 (PAH1), 10.49 (SIV) and 11.25$\upmu$m (PAH2). North is up and East is left in all images. The plots shown in the lower panel are the cuts made in all the filters at two different position angles. The left plot represents the emission in the torus (cut at PA=71$^{\circ}$) and the right plot represents the emission in the outflows (cut at PA=-23$^{\circ}$).}

\centering
\includegraphics[width=13cm]{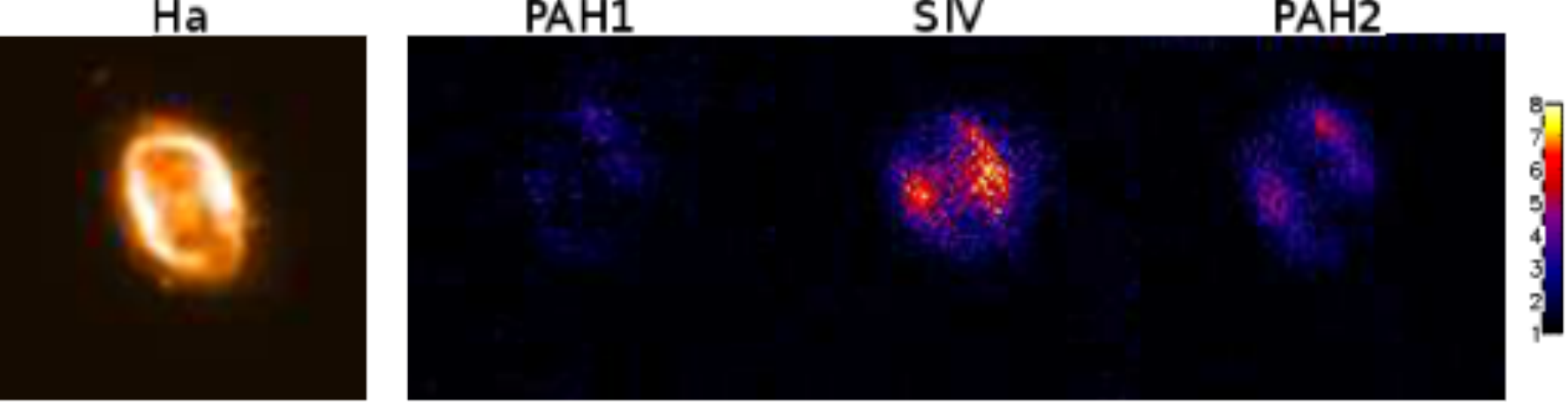}
\hbox{
\centering
\includegraphics[width=5.2cm]{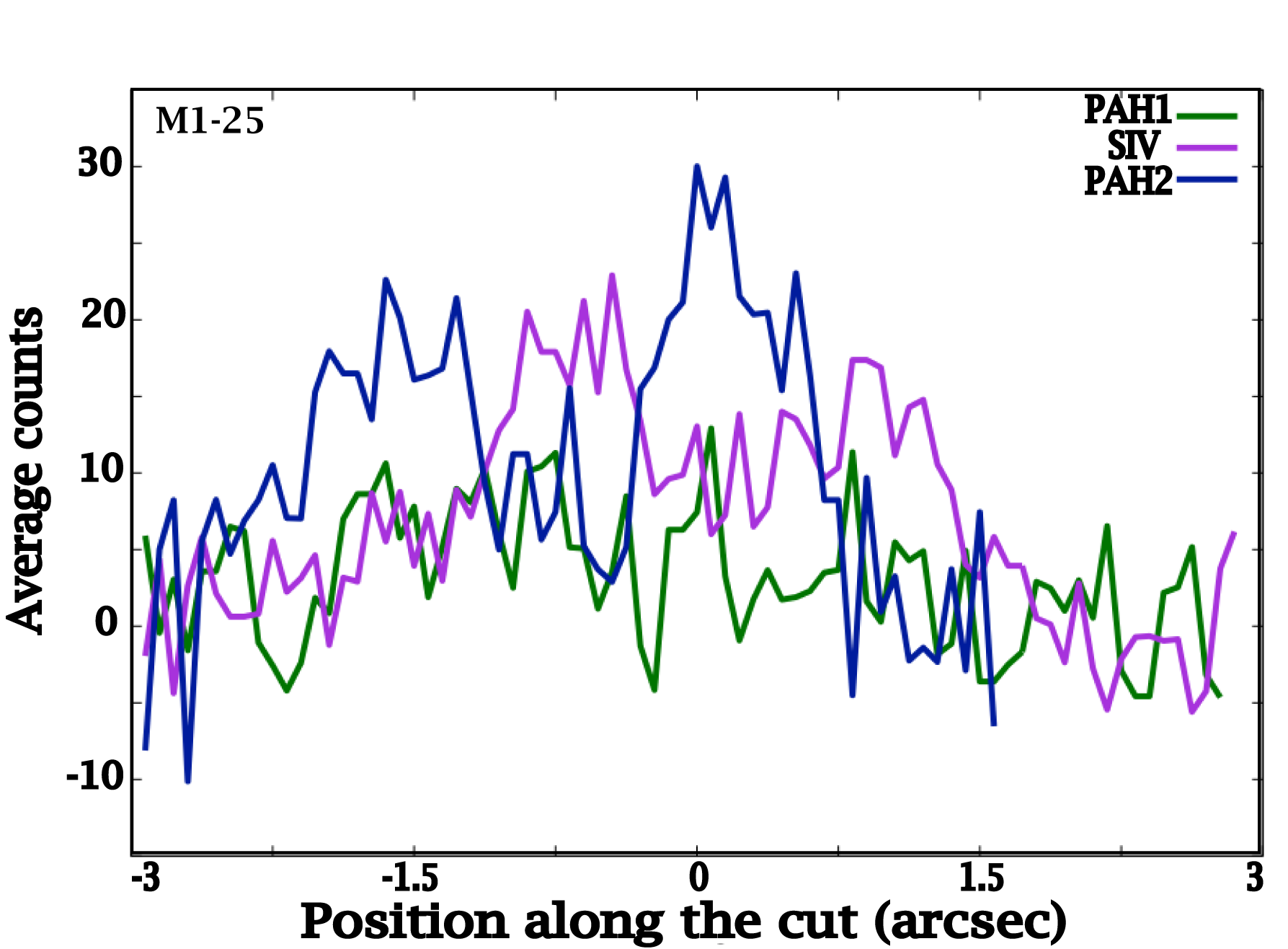}
\includegraphics[width=5.2cm]{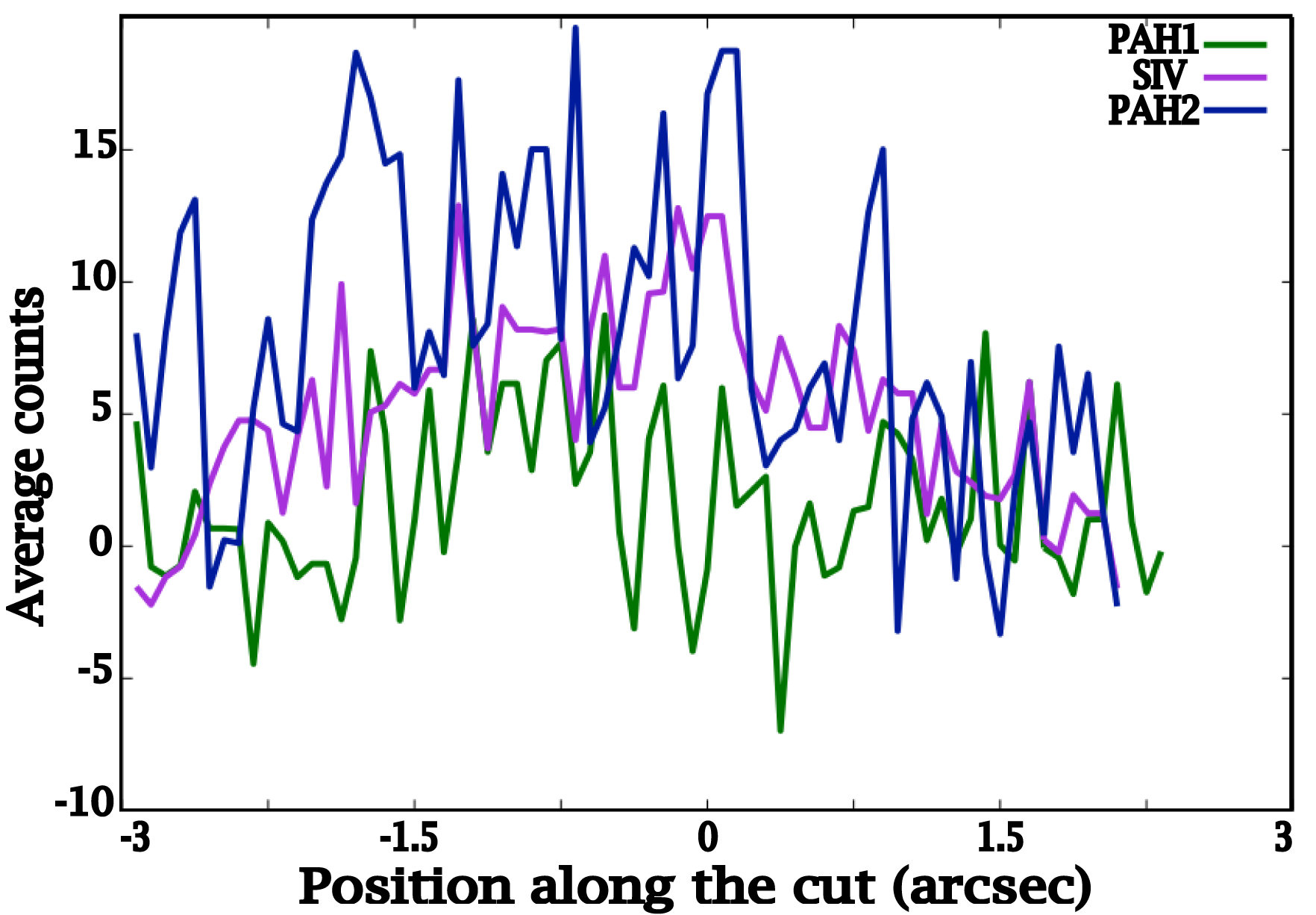}}
\caption[VISIR image of M1-25]{\label{m1-25}Same as Fig. \ref{cn1-5}, for M1-25.The left plot represents the emission in the torus (cut at PA=-66$^{\circ}$) and the right plot represents the emission in the outflows (cut at PA=32$^{\circ}$).}

 \centering
\includegraphics[width=7cm]{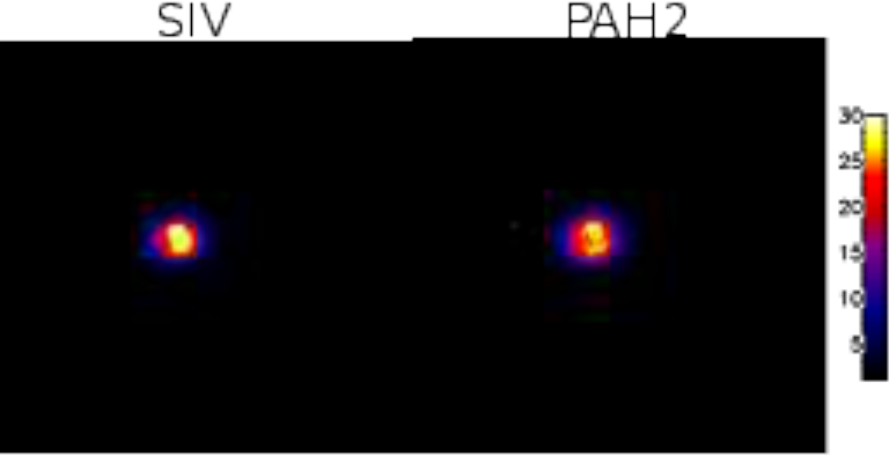}
\hbox{
\centering
\includegraphics[width=5.2cm]{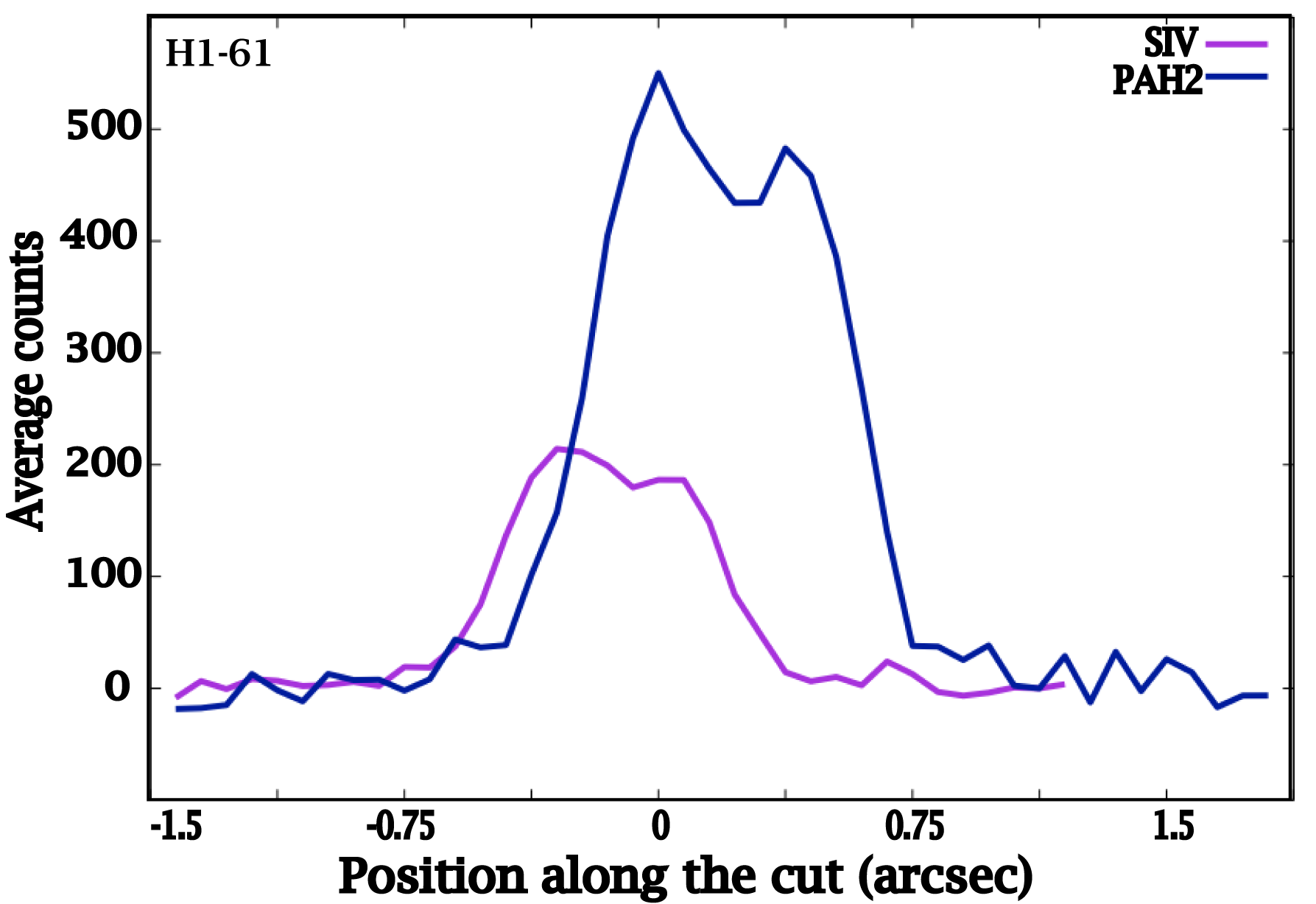}
\includegraphics[width=5.2cm]{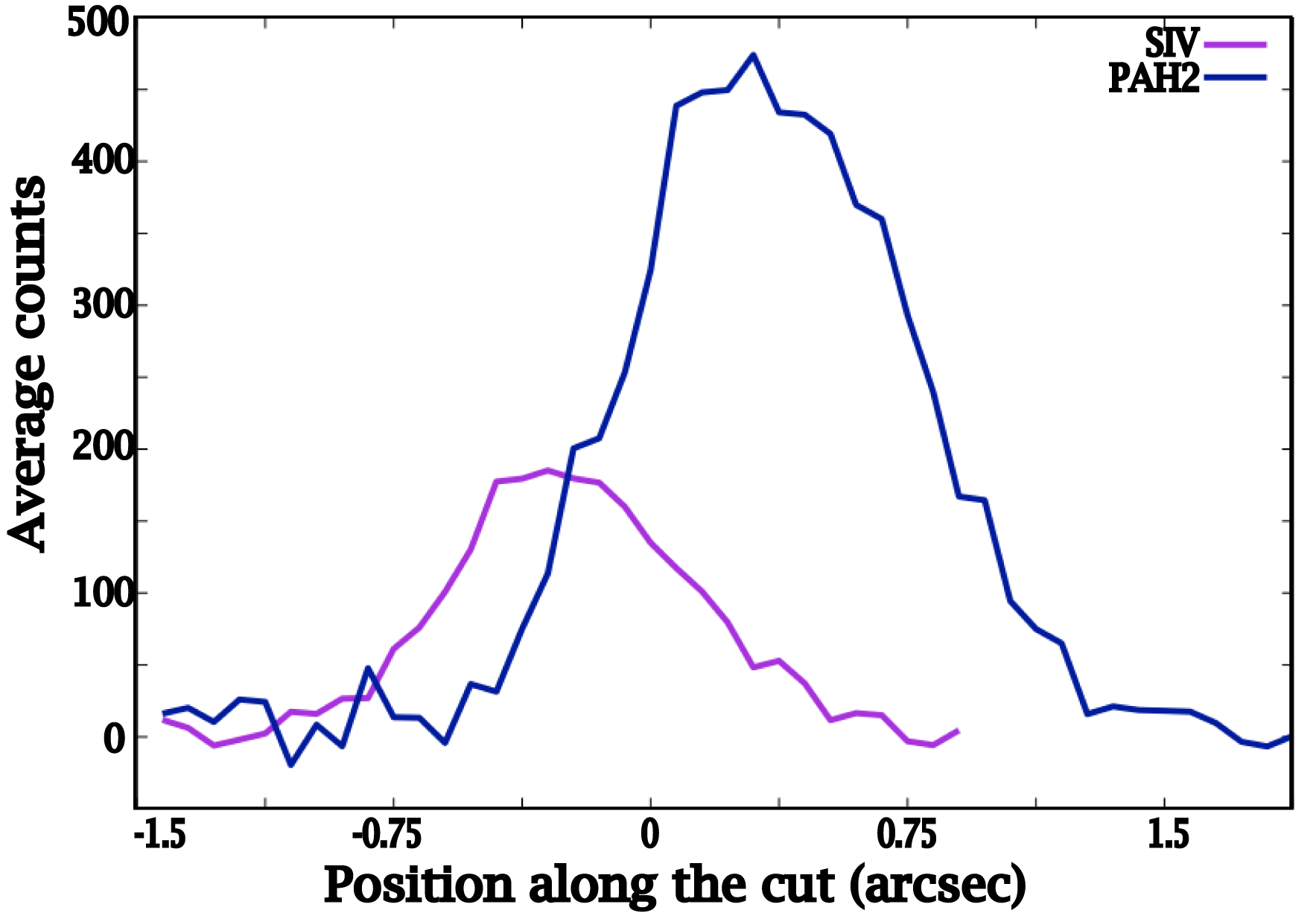}}
\caption[VISIR image of H1-61]{\label{h1-61}The leftmost panel displays the VISIR observations of H1-61 at 10.49 (SIV) and 11.25$\upmu$m
 (PAH2).  North is up and East is left. The rightmost plots show cuts made in all the filters at two different position angles. The left
 plot represents the emission in the torus (cut at PA=0$^{\circ}$), and the right plot represents the emission in the outflows (cut at PA=90$^{\circ}$).}
\end{figure*}

\begin{figure*}
 \centering
 \includegraphics[width=13cm]{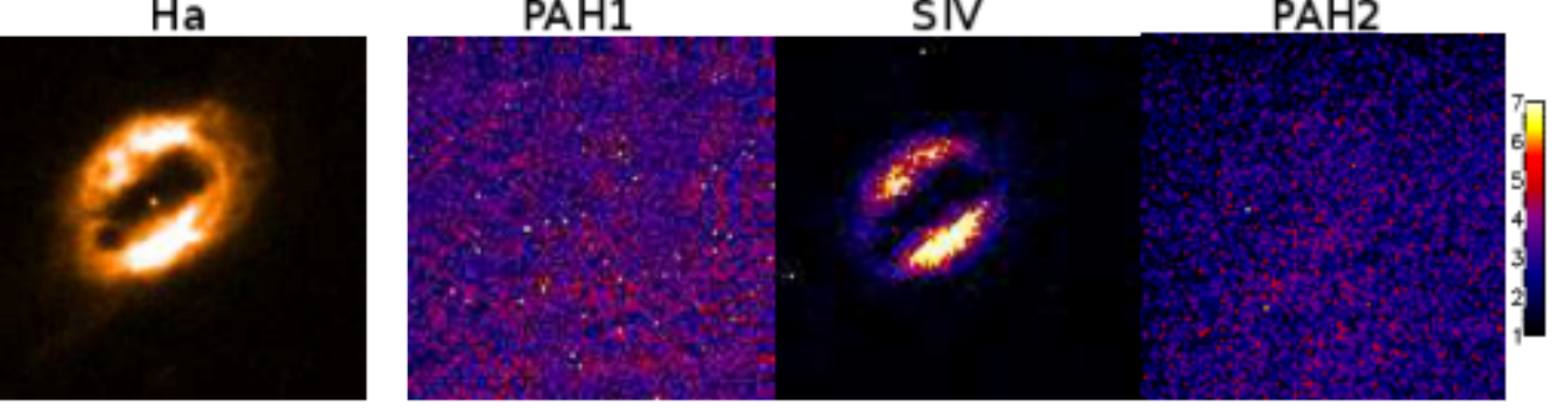}
 \hbox{
\centering
 \includegraphics[width=5.2cm]{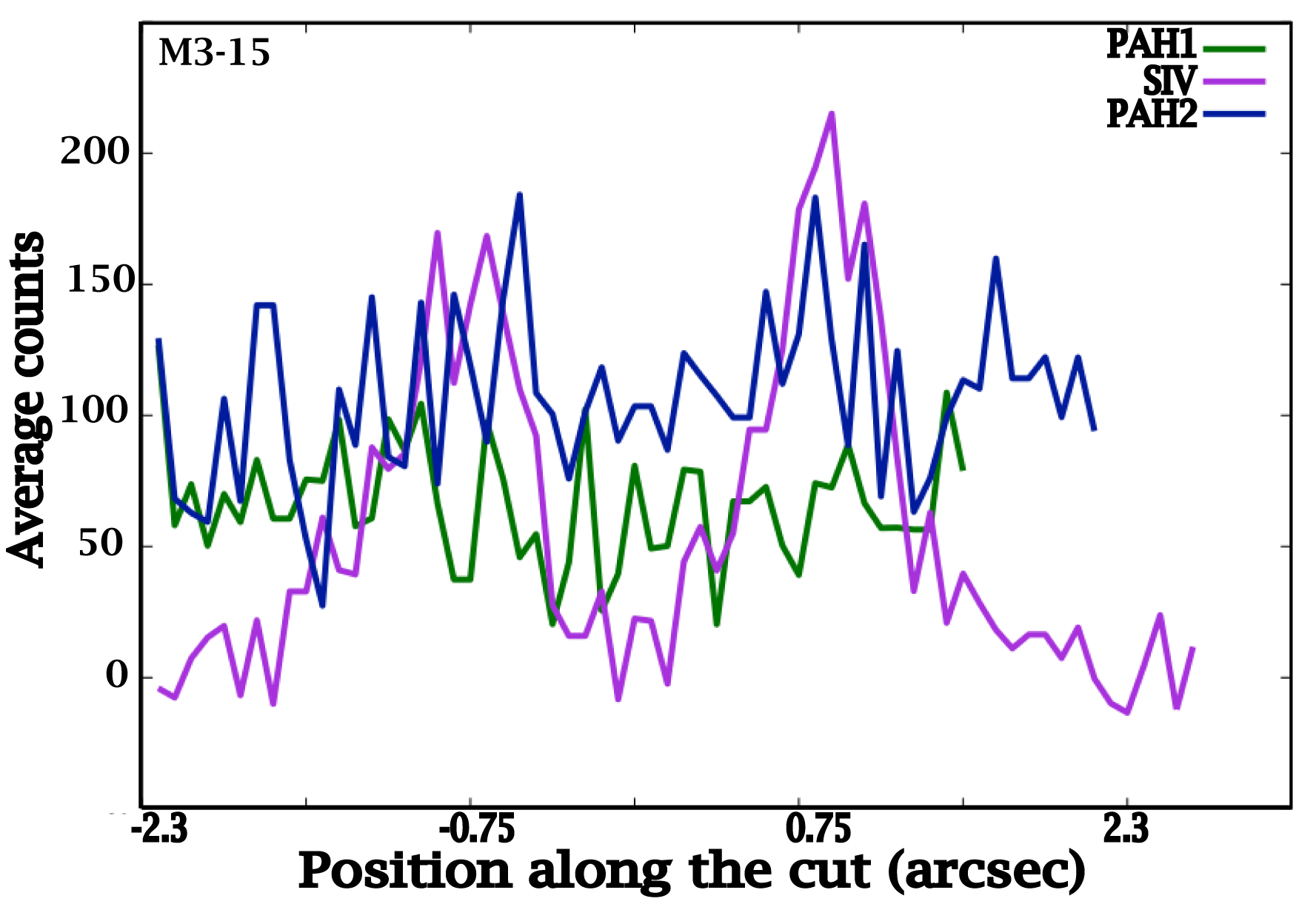}
}
\caption[VISIR image of M3-15]{ \label{m3-15}Same as Fig. \ref{cn1-5}, for M3-15. The plot represents the emission in the torus (cut at PA=34$^{\circ}$).}

\centering
\includegraphics[width=7cm]{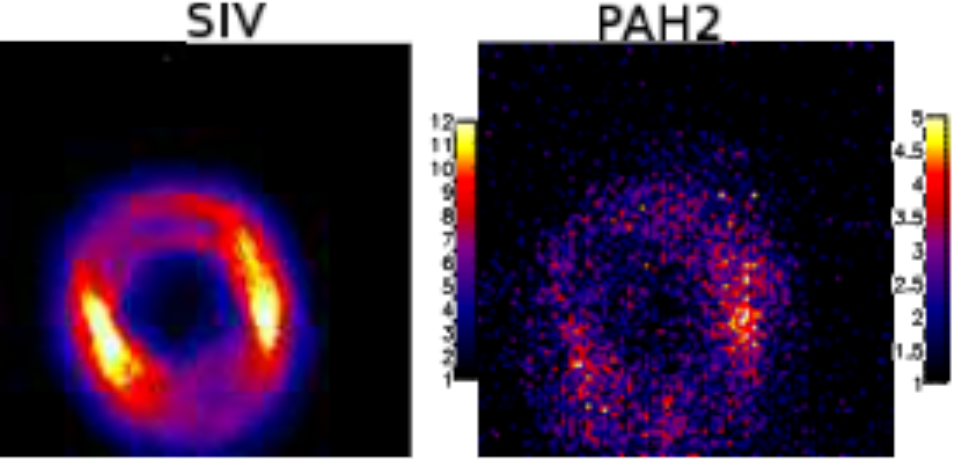}
\hbox{
\centering
\includegraphics[width=5.2cm]{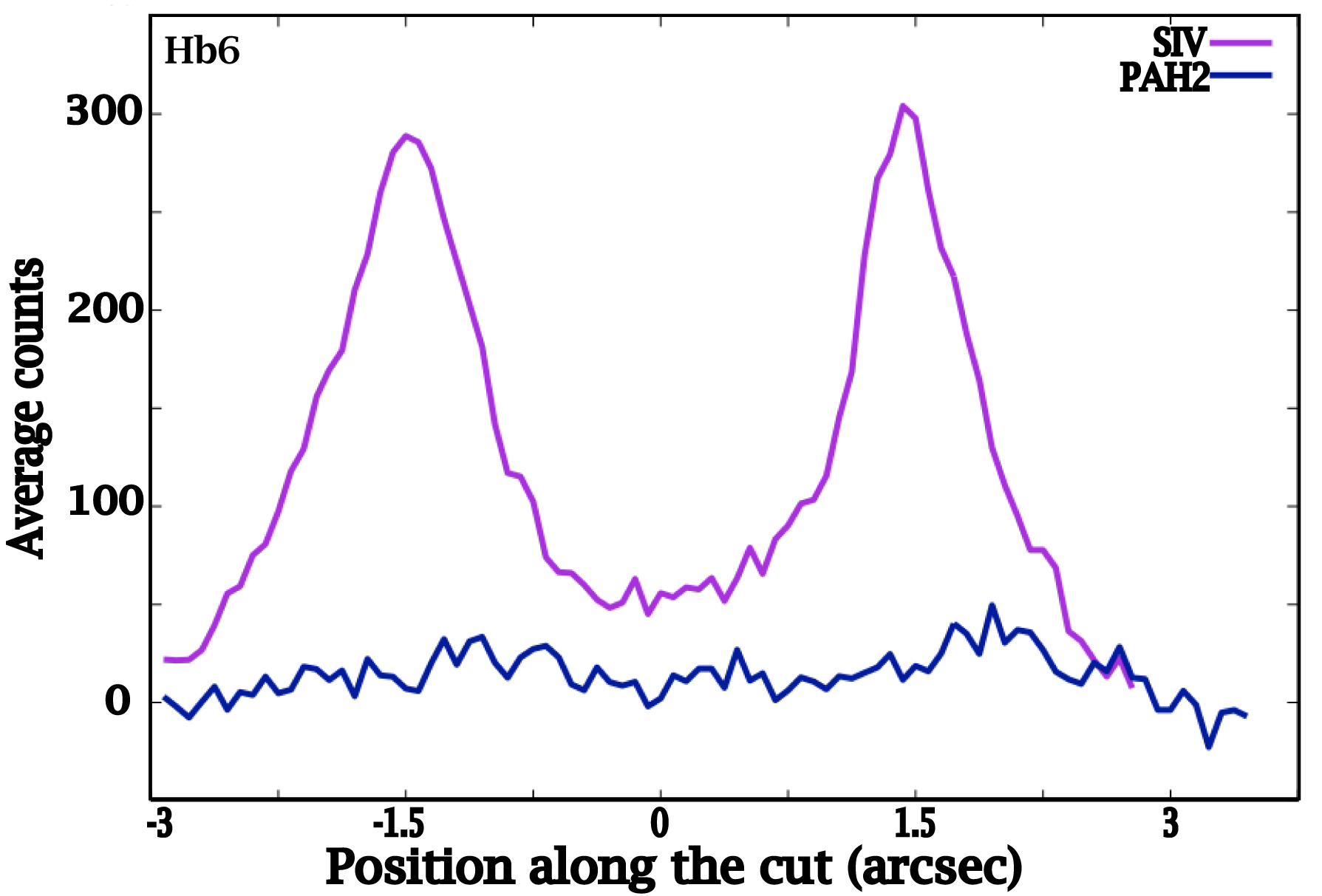}
\includegraphics[width=5.2cm]{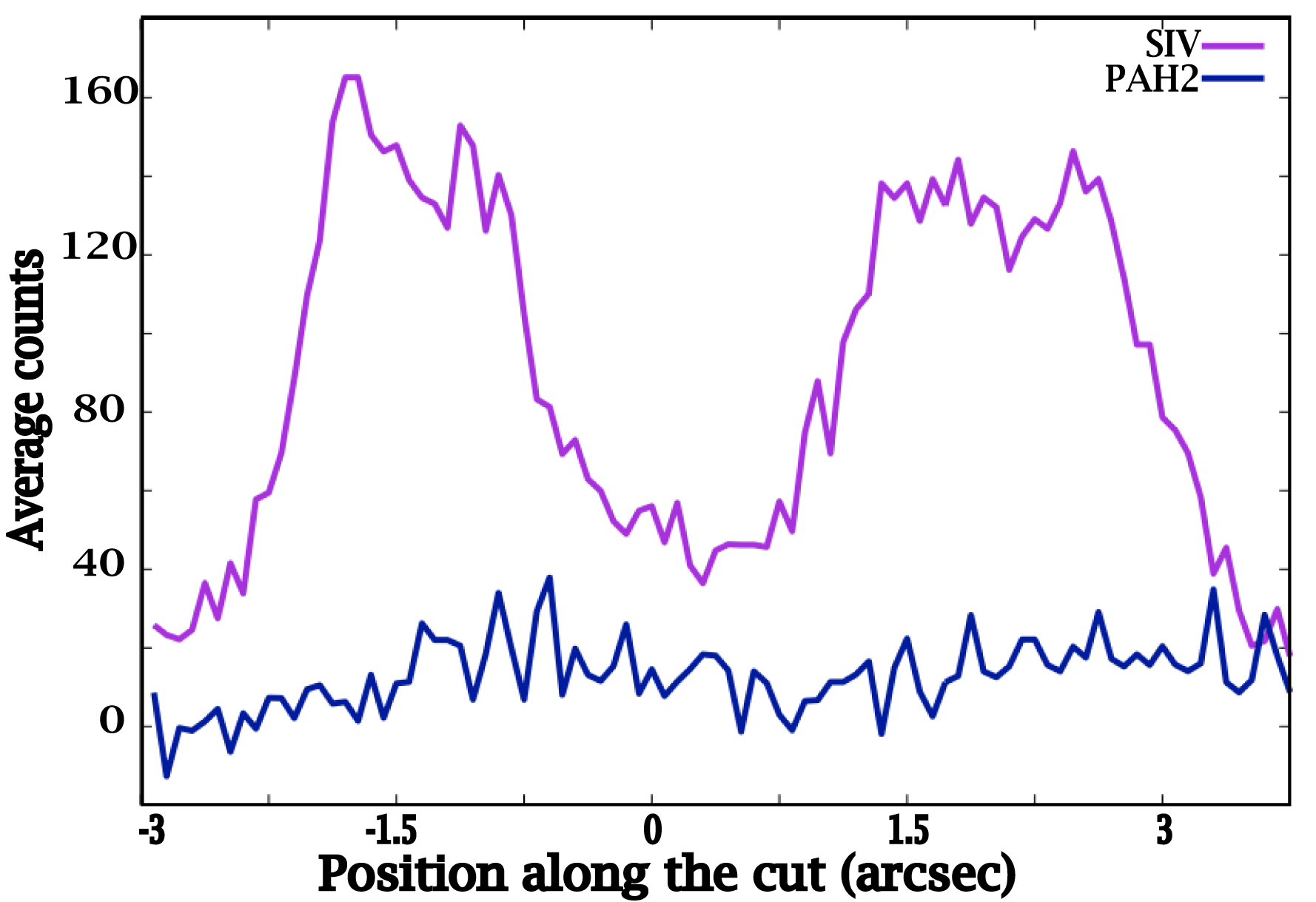}}
\caption[VISIR image of Hb6]{\label{hb6}Same as Fig. \ref{h1-61}, for Hb6. The left
 plot represents the emission in the torus (cut at PA=103$^{\circ}$), and the right plot represents the emission in the outflows (cut at PA=29$^{\circ}$).}

 \centering
\includegraphics[width=13cm]{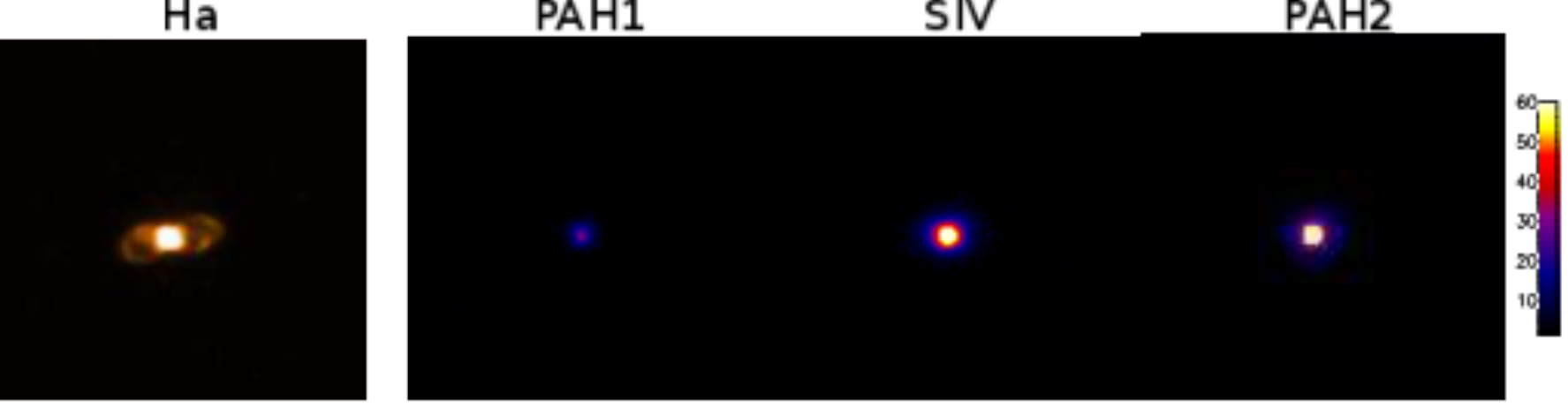}
\hbox{
\centering
 \includegraphics[width=5.2cm]{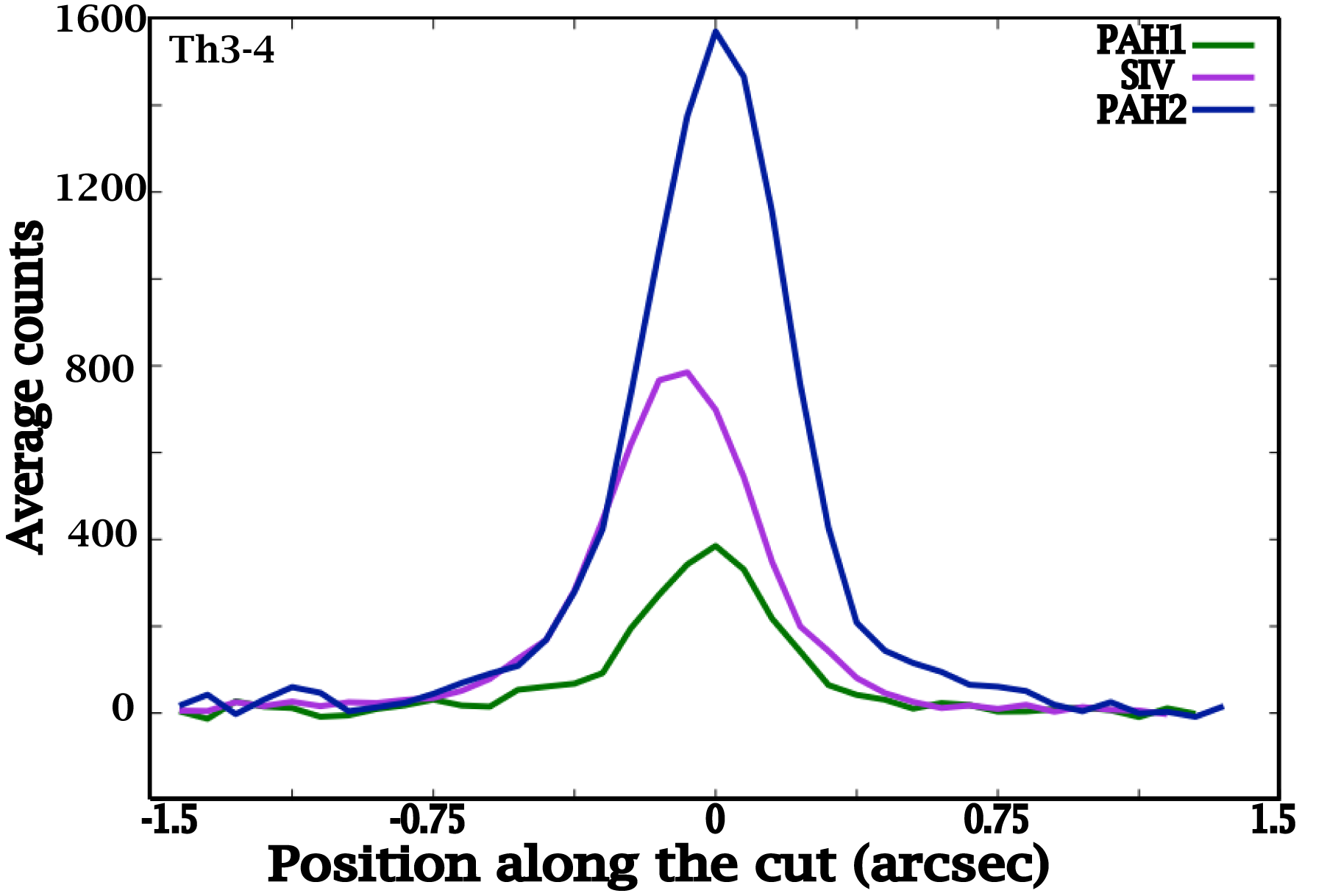}
\includegraphics[width=5.2cm]{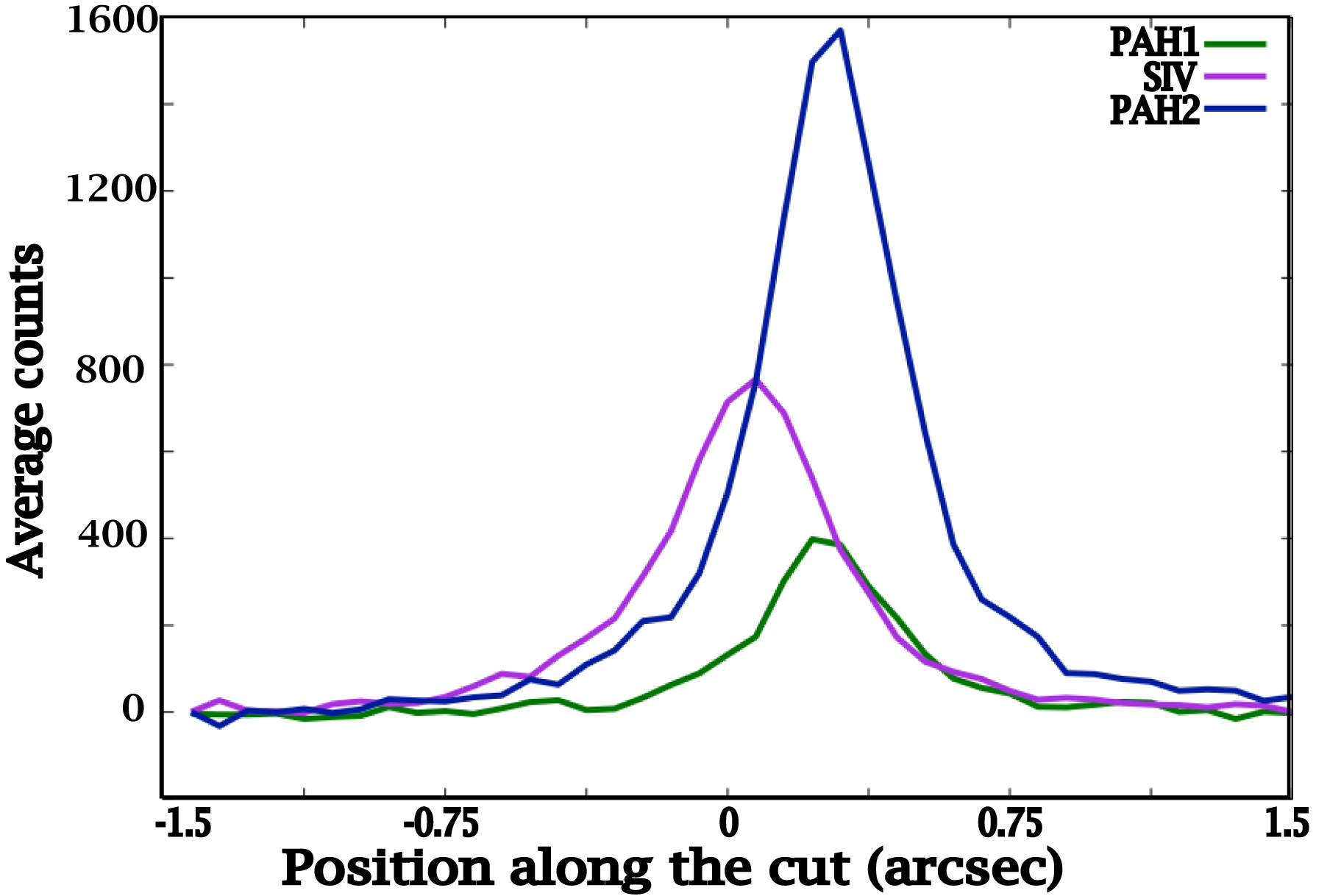}}
 \caption[VISIR image of Th3-4]{\label{th3-4}Same as Fig. \ref{cn1-5}, for Th3-4. The left
 plot represents the emission in the torus (cut at PA=4$^{\circ}$), and the right plot represents the emission in the outflows (cut at PA=95$^{\circ}$).}
\end{figure*}     

\begin{figure*}
 \centering
\includegraphics[width=13cm]{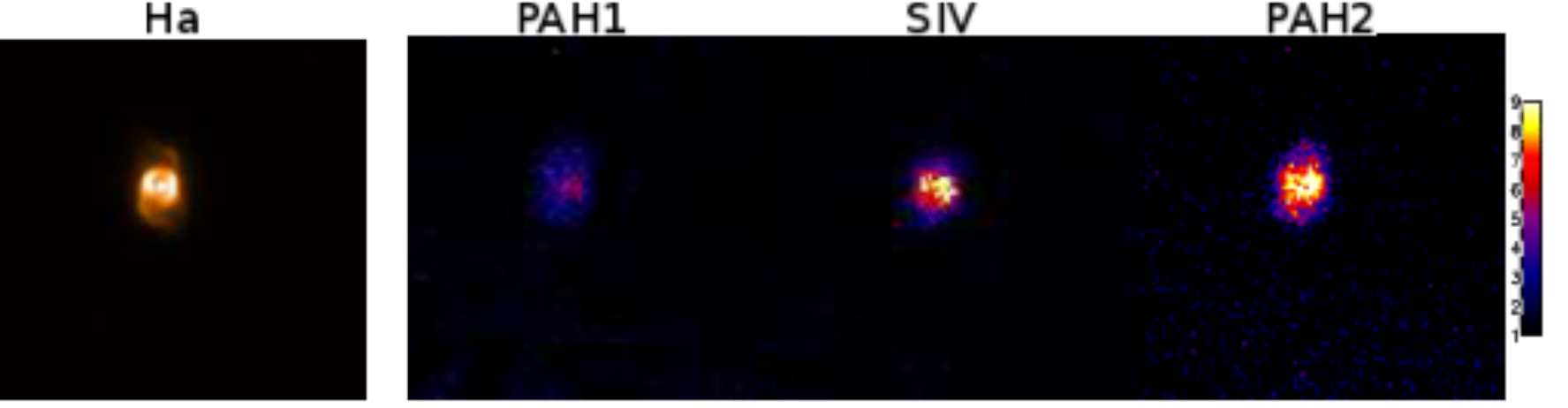}
\hbox{
\centering
 \includegraphics[width=5.2cm]{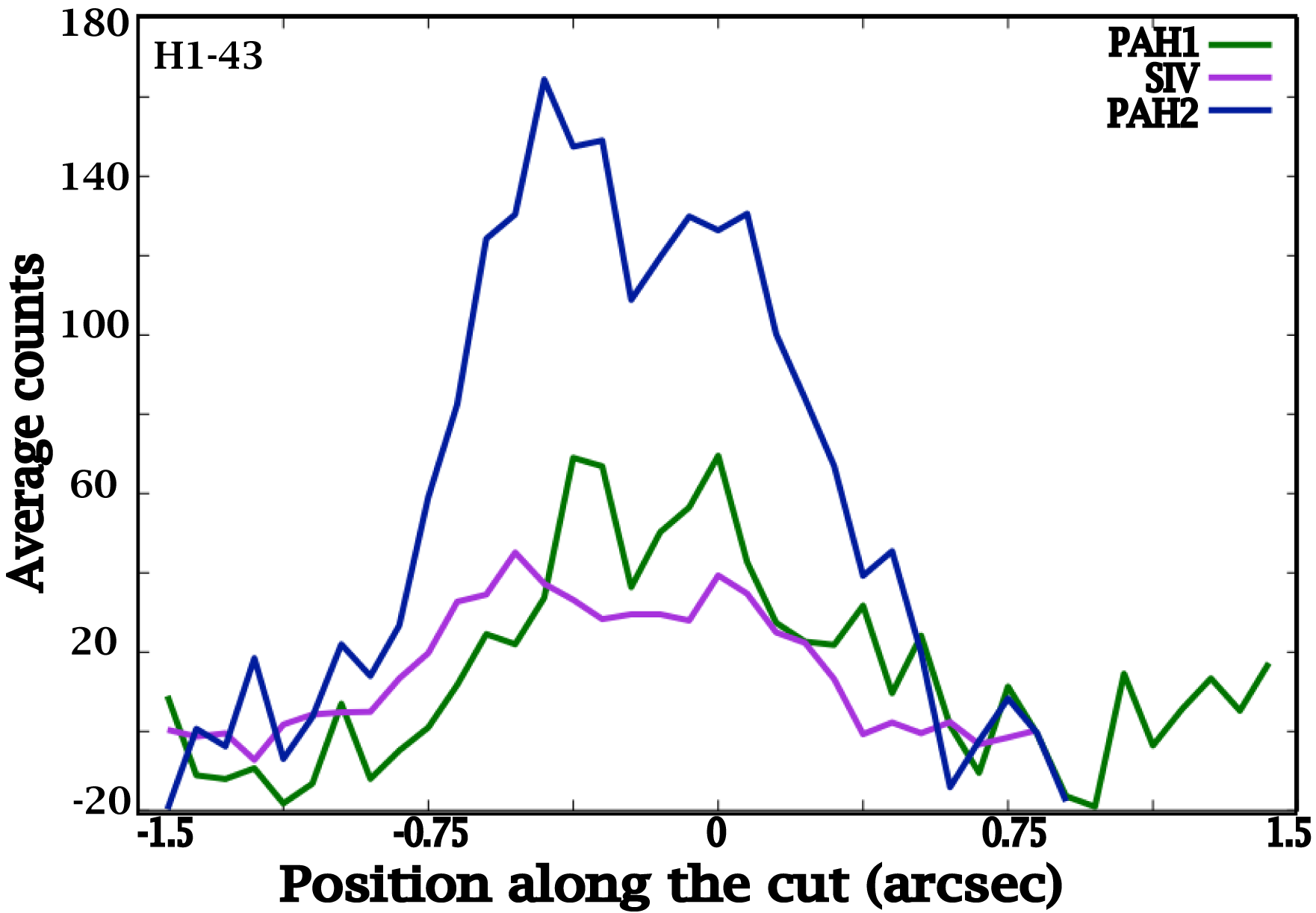}
\includegraphics[width=5.2cm]{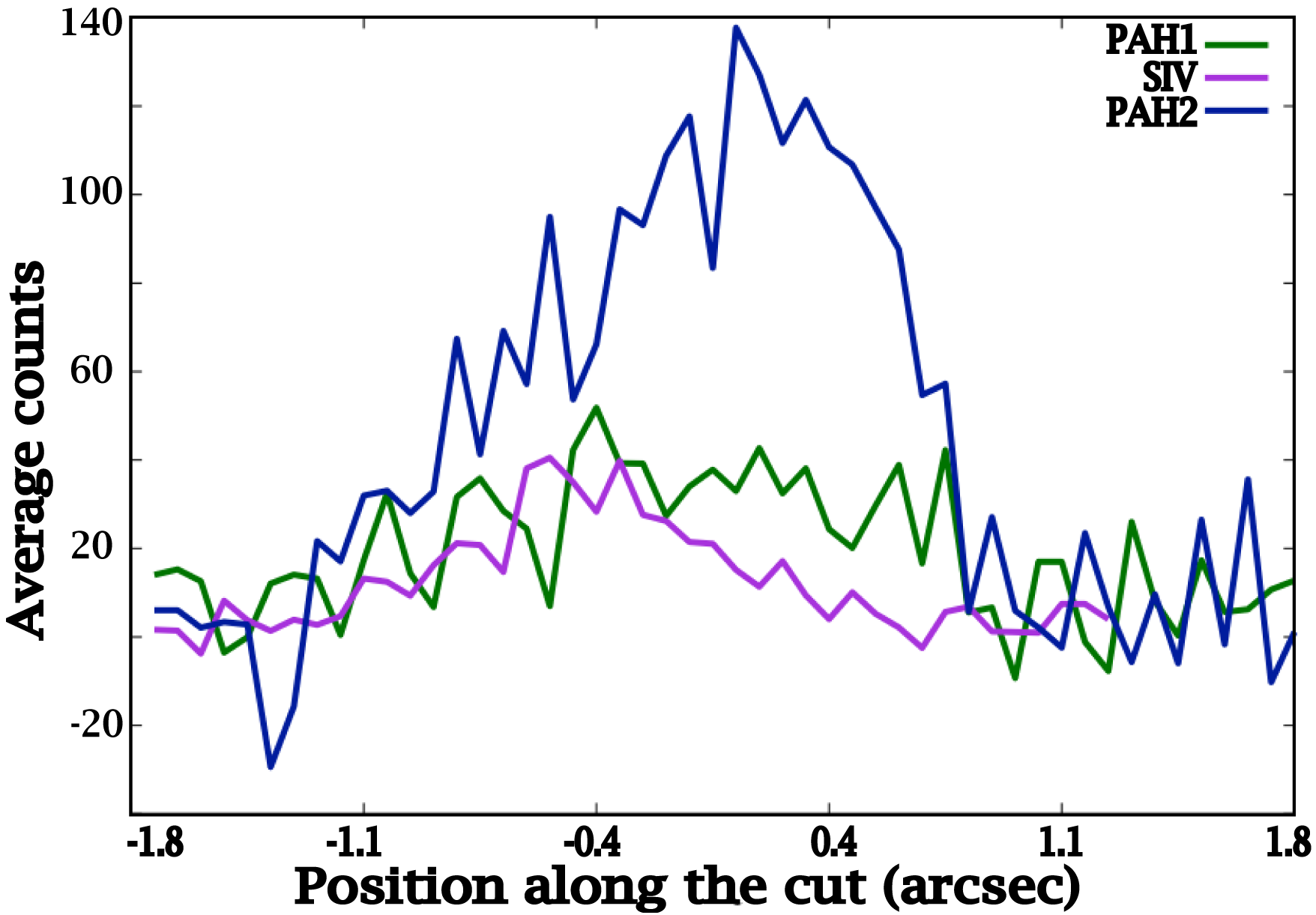}}
\caption[VISIR image of H1-23]{\label{h1-43}Same as Fig. \ref{cn1-5}, for H1-43. The left
 plot represents the emission in the torus (cut at PA=80$^{\circ}$), and the right plot represents the emission in the outflows (cut at PA=-23$^{\circ}$).}

 \centering
\includegraphics[width=10.75cm]{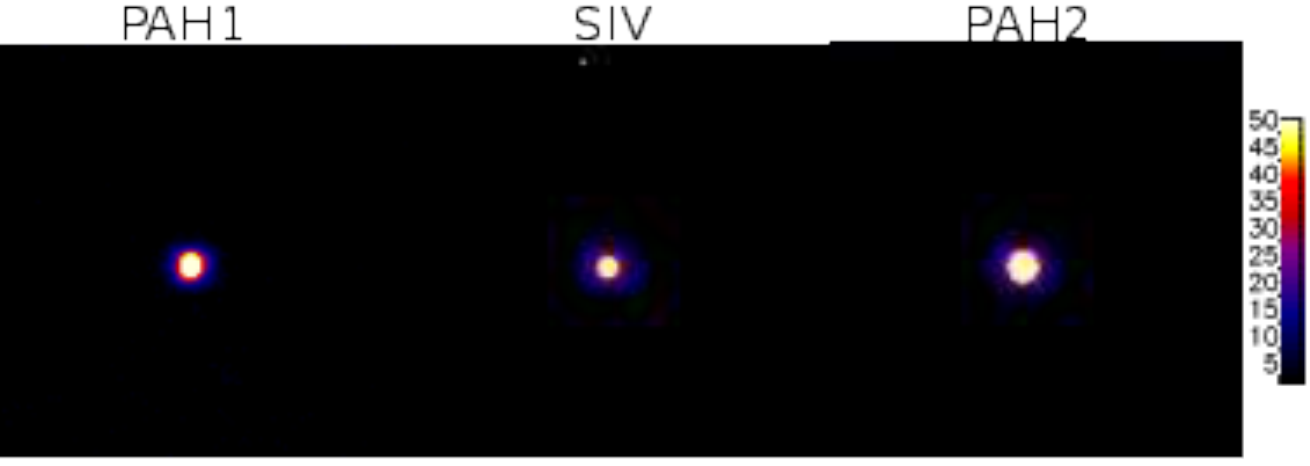}
\hbox{
\centering
 \includegraphics[width=5.2cm]{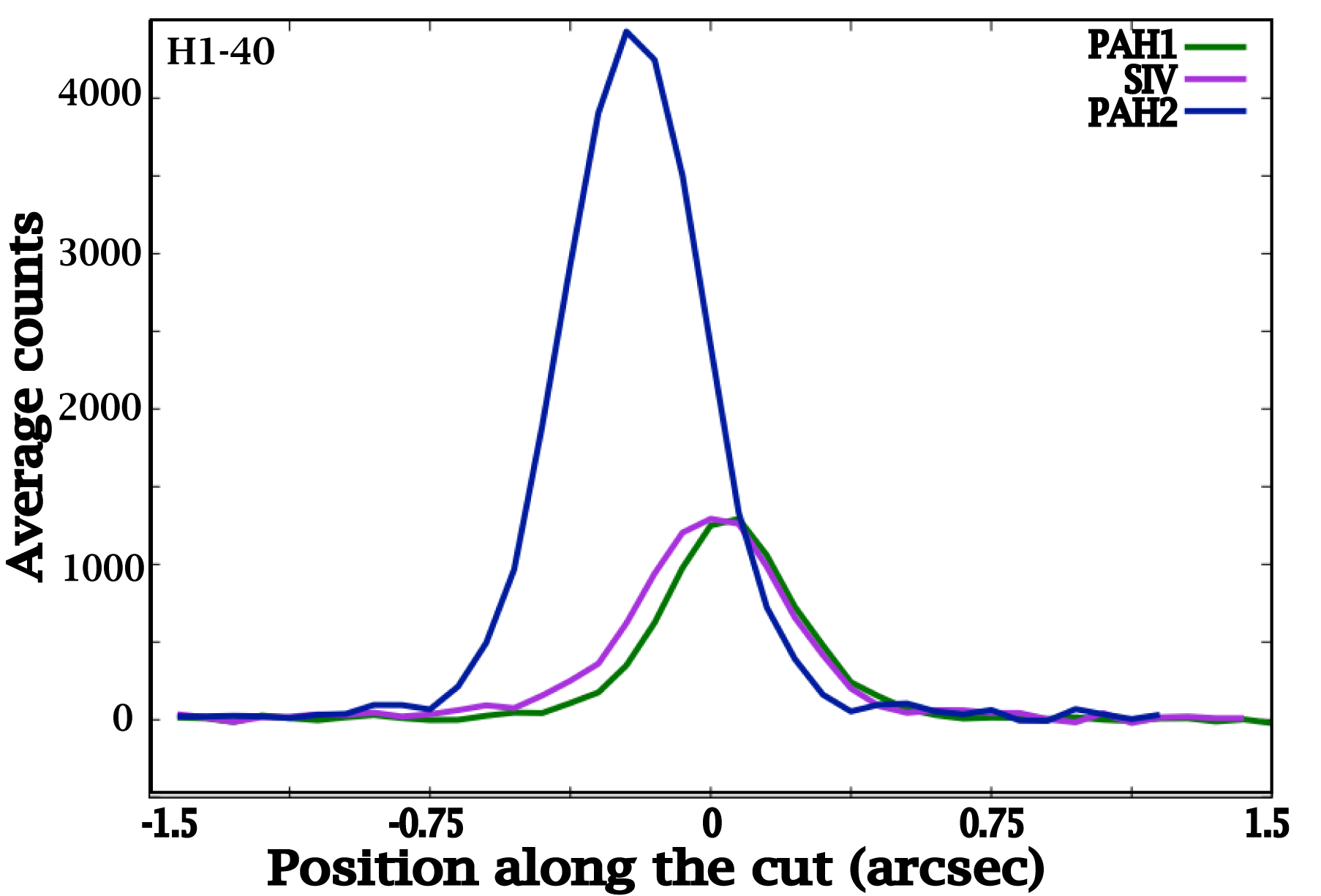}
\includegraphics[width=5.2cm]{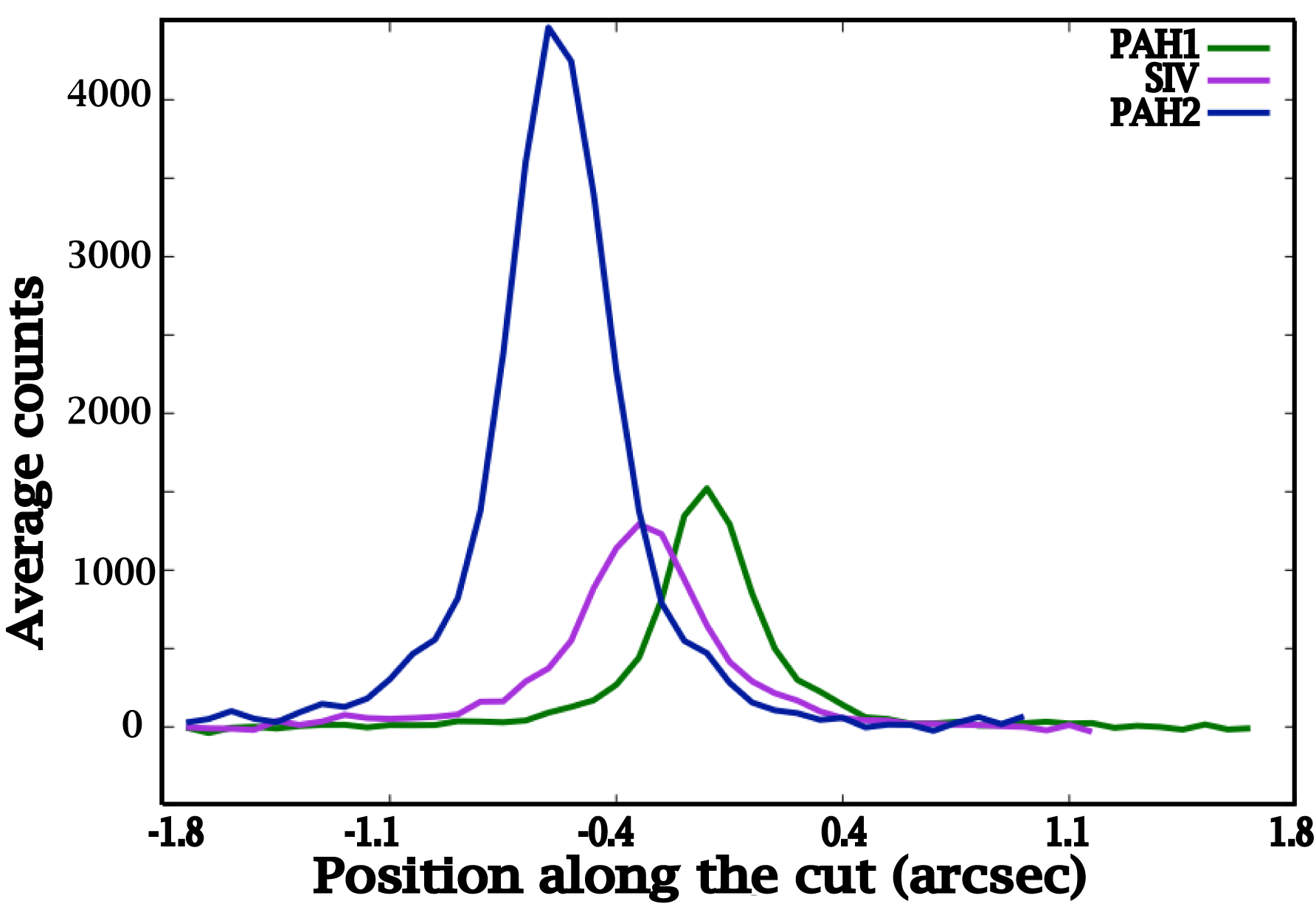}}
\caption[VISIR image of H1-40]{\label{h1-40}PN H1-40. VISIR observations at 8.59 (PAH1), 10.49 (SIV) and 11.25$\upmu$m (PAH2). North is 
up and East is left. The rightmost plots show the cuts made in all the filters at two different position angles. The left plot
 represents the emission in the torus (cut at PA=-4$^{\circ}$), and the right plot represents the emission in the outflows (cut at PA=83$^{\circ}$).}
\end{figure*}     

\subsection{VISIR images} 

In this Section, we present the imaging data from those PNe where no spectroscopy was acquired before presenting 
the data from the three objects where a combination of imaging and spectroscopy was obtained. Where possible, we use 
{\it HST} images to obtain further information about their morphology. For the PNe Cn1-5 and M1-25 the {\it HST} images
 were taken from \citet{rees13}. For M3-15, Th3-4, H1-43, M1-31, and M3-38 the {\it HST} images were taken 
from \citet{me11}. All the images, shown in Figs 1 to 11, have exactly the same field of view (FOV) of
 7\arcsec$\times$ 7\arcsec{}, including the {\it HST} images (unless otherwise stated in the caption, Fig 1). 
The scale bar next to the VISIR images represents the rms scale. 
As most of our objects are red, meaning that they are brighter at longer wavelengths, the PAH1 images are very
 noisy or the object is barely detected, except for four compact sources. 

 For a better understanding of the distribution of the emission on each filter we make two different cuts across 
the images.  In each figure, we display these two cuts (one made along the torus and a second along the outflows)
 in order to investigate the relative strength and distribution of PAHs and ionised material in these directions.
 
{\bf Cn1-5} (PN G002.2-09.4, Fig. \ref{cn1-5}): Here, we will use the {\it HST} image to define torus and outflows in this object.
 The ring-like structure with the central star in the middle, is the torus. The fact that we can see the edges brighter is a projection effect. This structure have also been seen in HaTr4 (\citealp{tyndall12}). The outflows would 
then, therefore, be roughly perpendicular to the torus. In the VISIR images, we can see very faint emission on the PAH1 filter. For
 the SIV filter, the emission is very bright at the edges of the torus, correlating with the brightest parts in the H$\alpha$ image.
 In the case of the PAH2 filter, the PAHs seem to be also in the torus, but in a more concentrated region than the SIV emission. The 
two plots below the VISIR images display cuts made along the torus (left) and along the outflows (right) at a PAs of 71$^{\circ}$ and
 -23$^{\circ}$, respectively (the cuts are not made perpendicularly in order to account for the non-axisymmetric nature of the object). 
The plots show that the PAHs at 11.3$\upmu$m are very bright in the torus, and with a double-peaked structure, due to the two sides of
 the torus. Also, the peak of the PAHs seems to be  slightly offset compared to the peak of the SIV emission. For the second plot, showing
 the brightness distributions along the outflows, the emission shows little structure and similar level for all the three filters.

{\bf M1-25} (PN G004.9 +04.9, Fig. \ref{m1-25}): The H$\alpha$ {\it HST} image shows an elongated ring with two brightened opposed edges. 
This elongated (almost face-on) ring is the torus. The outflows in this case should be perpendicular to the torus, along the major axis of the nebula. In the VISIR images, again the emission in the PAH1 filter is very faint. In the SIV filter, the two bright emission peaks 
are a projection of the dusty central torus seen in the 
 H$\alpha$ image and also resolved
by VISIR. In the case of the PAH2 filter, we can see that the structure looks 
similar to what is seen  in the SIV filter, but with an offset between the brightest parts in both filters. This
 is also evident in the plot  showing the cut through the torus. It shows the typical two-peaked structure of the torus, but  the peak 
of the SIV filter is exactly in between the two peaks of the PAH2 filter. This plot was made doing a cut on the three images along a PA=-66$^{\circ}$ 
along the brightest parts of the torus. In the rightmost plot we present a cut along the outflows. As the PN is almost face-on, the
 outflows are blended with the torus. The cut was made at a PA of 32$^{\circ}$  is not very informative about the emission distribution along the poles.

{\bf H1-61} (PN G006.5 --03.1, Fig. \ref{h1-61}): VISIR imaging shows compact emission in both SIV and PAH2 filters. While the object is
 barely resolved by the instrument, in the PAH2 filter, we can still see that the emission displays a double-peaked structure. This 
structure is shown more clearly by the cut along the torus (at a PA of 0$^\circ$), while the cut through the outflows (perpendicular 
to the torus at a PA of 90$^\circ$) shows a single peak with an offset between the centroid in the SIV and PAH2 filters.

{\bf M3-15} (PN G006.8+04.1, Fig. \ref{m3-15}): In the H$\alpha$ image, we can see an elongated ring-like structure with two very bright edges.
 These two bright edges are indicative of an inclined ring-like structure, where a projection effect is responsible for making the edges appear
 both brighter and broader. This structure is the torus and similarly to the PN M1-25, the outflows are certainly  perpendicular to this structure, 
and so orientated almost along the line of sight. As for the VISIR images, this PN was only detected in the SIV filter. Its morphology in this band 
leads us to dub this object the {\it Kiss Nebula}. It was not detected in either PAH filters, although, we know that this PN does show PAH emission as
 they were detected using {\it Spitzer} \citep{me11}. The brightest emission in the SIV filter correlates very well with he brightest emission in 
the H$\alpha$ image. The plot below the VISIR images, represent a cut along the torus at a PA of 34$^{\circ}$, the SIV emission is double-peaked 
and there is no emission detected in the PAHs filters. For this PN we do not show a cut along the outflows cause they seem to be in the line-of-sight. 

{\bf Hb6} (PN G007.2+01.8, Fig. \ref{hb6}): In the SIV filter image, we can see a very well-defined, face-on ring, 
which splits into two faint arcs perpendicular to the axis of the brightened regions. In the PAH2 filter image, 
 we can see faint PAH emission which follows the same ring-like structure. The two plots next to the VISIR images,
 represent cuts along two different angles. For the plot on the left, the cut was made at a PA of 103$^{\circ}$. 
This cut goes along the two brightest points. In the plot we can see that [SIV] emission is much brighter than 
the PAH emission in the PAH2 filter, and it also shows the double-peaked structure distinctive of a toroidal
 morphology. The rightmost plot shows a cut made at a PA of 29$^{\circ}$, almost perpendicular to the first
 cut (again chosen not to be exactly perpendicular in order to account for non-axisymmetry in the object). In
 this plot we can see that the SIV filter shows the double-peaked structure, but also it shows sub-structure 
that matches the faint double ring in the image. As for the PAH2 filter, the emission is too faint and shows no discernible structure.     

{\bf Th3-4} (PN G354.5+03.3, Fig. \ref{th3-4}): The H$\alpha$ {\it HST} image shows a very compact central 
nebula  with two opposing outflows orientated roughly East-West.
In the VISIR imaging, the object is barely resolved. The PAH1 filter image shows a very faint detection, while
 through the SIV and PAH2 filters we detect a faint compact emission associated with the central nebula as visible 
in the H$\alpha$ image. Cuts were made at PAs of 95$^\circ$ and 4$^\circ$ through the outflows and torus, respectively.
 In both cuts, the emission in all filters shows a single peak, corresponding the to central nebular region and no
 further discernible structure that can be associated to the outflows or torus.

{\bf H1-43} (PN G357.1 -04.7, Fig. \ref{h1-43}): The H$\alpha$ image shows a ring-like structure
 surrounding the central star with faint outflows projecting to the North and South of this central
 region. In the VISIR images, we can see a very faint detection in the PAH1 filter. The emission in
 the SIV filter is concentrated in the inner parts of the nebula as delineated by the torus visible
 in the H$\alpha$ image while still displaying a double-peaked profile (as shown in the cut through 
the torus taken at a PA of 80$^\circ$).  The PAH2 filter image shows emission over a more extended 
region particularly along the direction of the outflows seen in the H$\alpha$ image.  The emission profile in all
 filters is double-peaked in a cut through the toroidal structure, but while the PAH emission occurs
 over a more extended region its peaks are found inside those of the [SIV] emission.  A cut through
 the outflows (at a PA of -23$^\circ$) indicates that little structure is detected in the PAH1 and
 SIV filters, while the PAH2 image shows a broad off-centre emission towards the North-west of the central star (also clear from inspecting the image).

{\bf H1-40} (PN G359.7-02.6, Fig. \ref{h1-40})  The VISIR imagery shows a morphology very similar 
to that of Th3-4, with a compact, central dusty structure. The PAH1 and PAH2 filter images show 
slightly more extended emission than that with the SIV filter, but still with no clearly discernible 
structure.  Cuts taken to pass through the major (PA of -4$^\circ$) and minor (PA of 83$^\circ$) axis 
of this extended emission (corresponding to the torus and outflows, respectively) both show a similar 
compact single-peaked profile in all filters. It is interesting to mention that the peak of the emission
 in the PAH1 filter and the SIV filter are very well matched in position for the cut through the torus, 
but offset from the peak of the PAH2 peak emission (contrary to that observed for Th3-4). For the cut
through the outflows, we can see that the peak of emission in all three filters seem to be offset.

\subsection{Objects with VISIR spectroscopy}

\begin{figure*}
 \centering
     \vbox{
     \includegraphics[width=15.5cm]{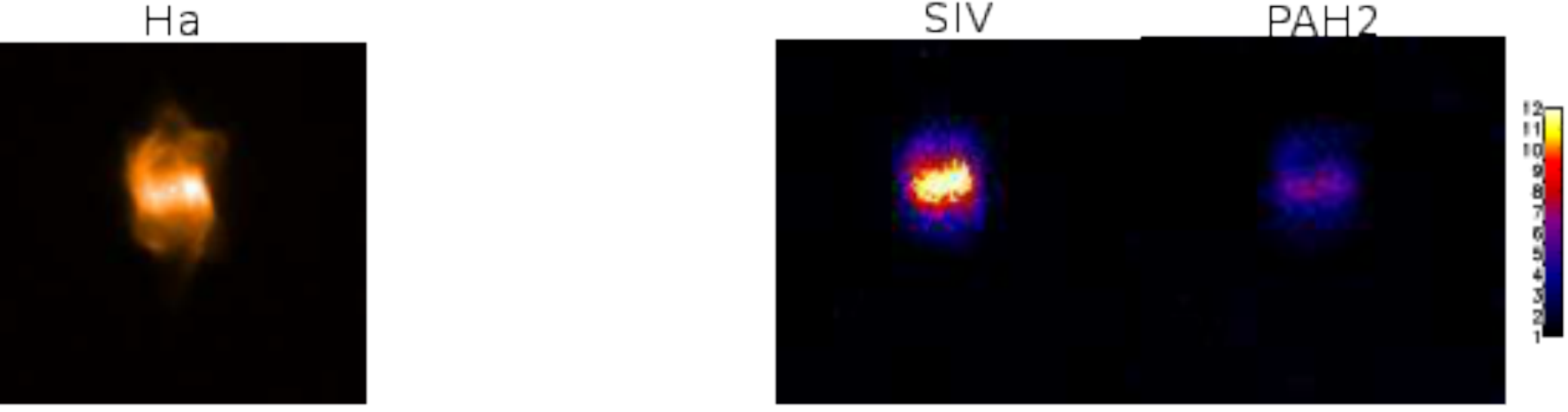}
     \includegraphics[width=5.7cm,height=3.85cm]{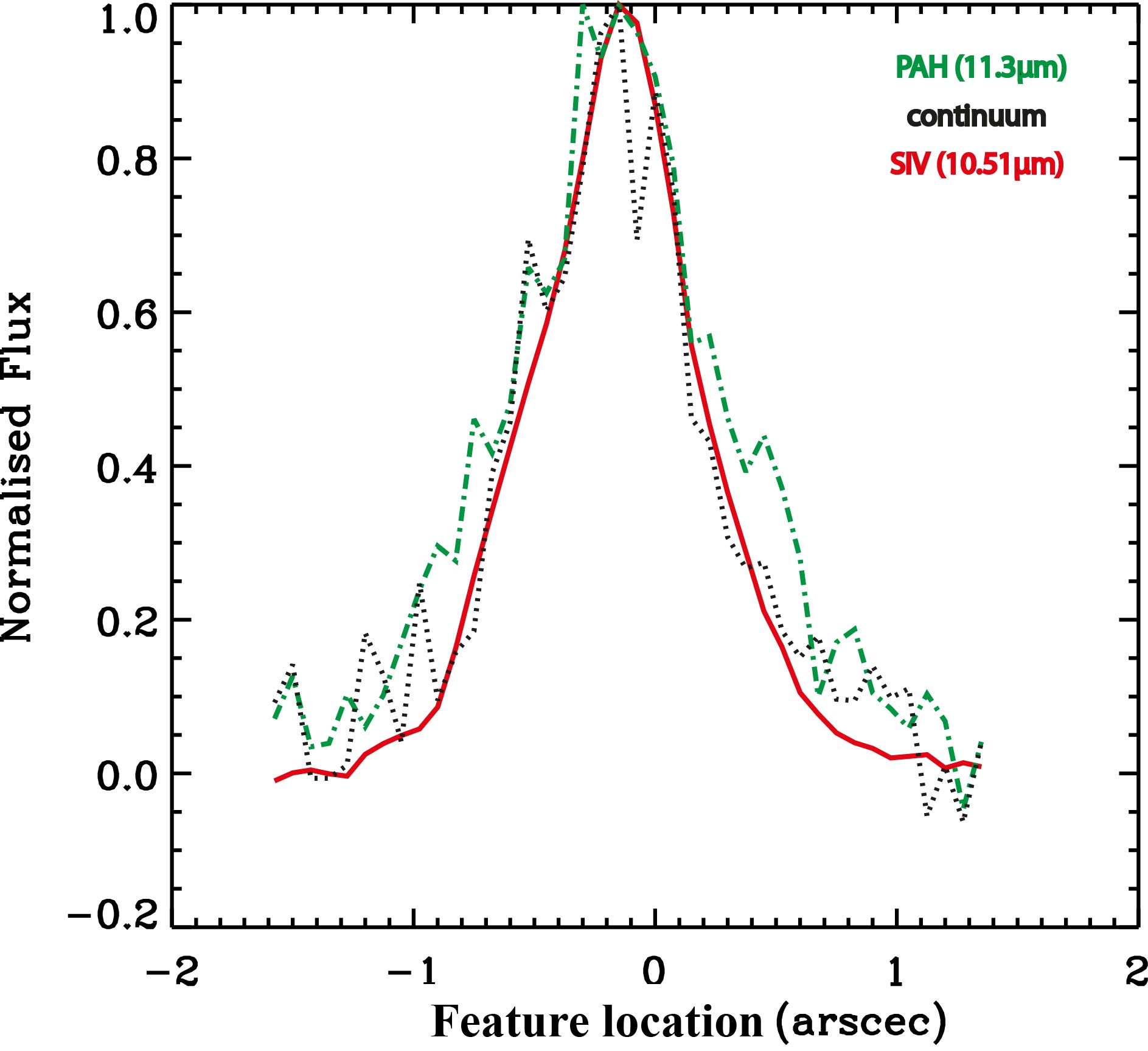}
     \includegraphics[width=5.7cm]{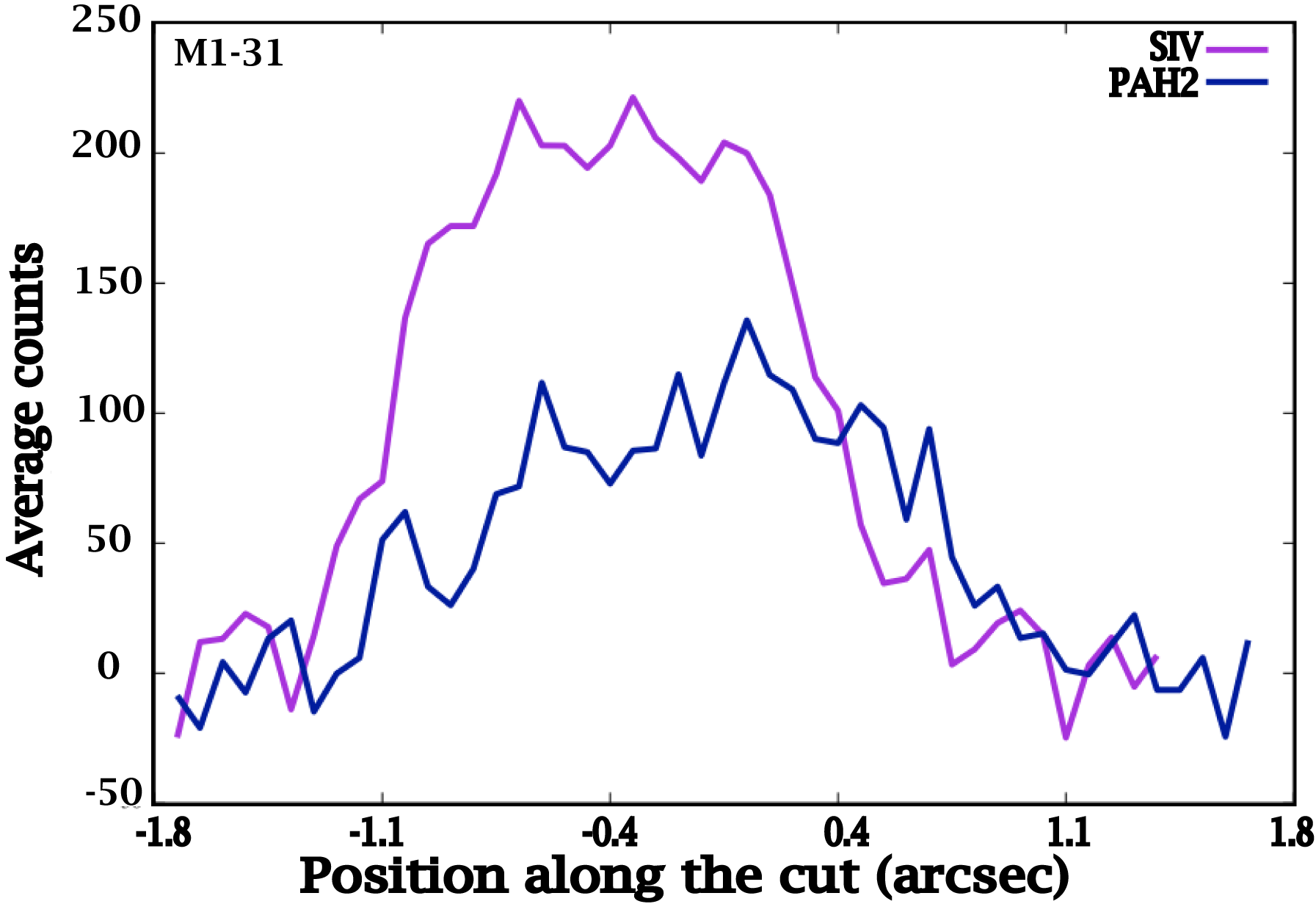}
\includegraphics[width=5.7cm]{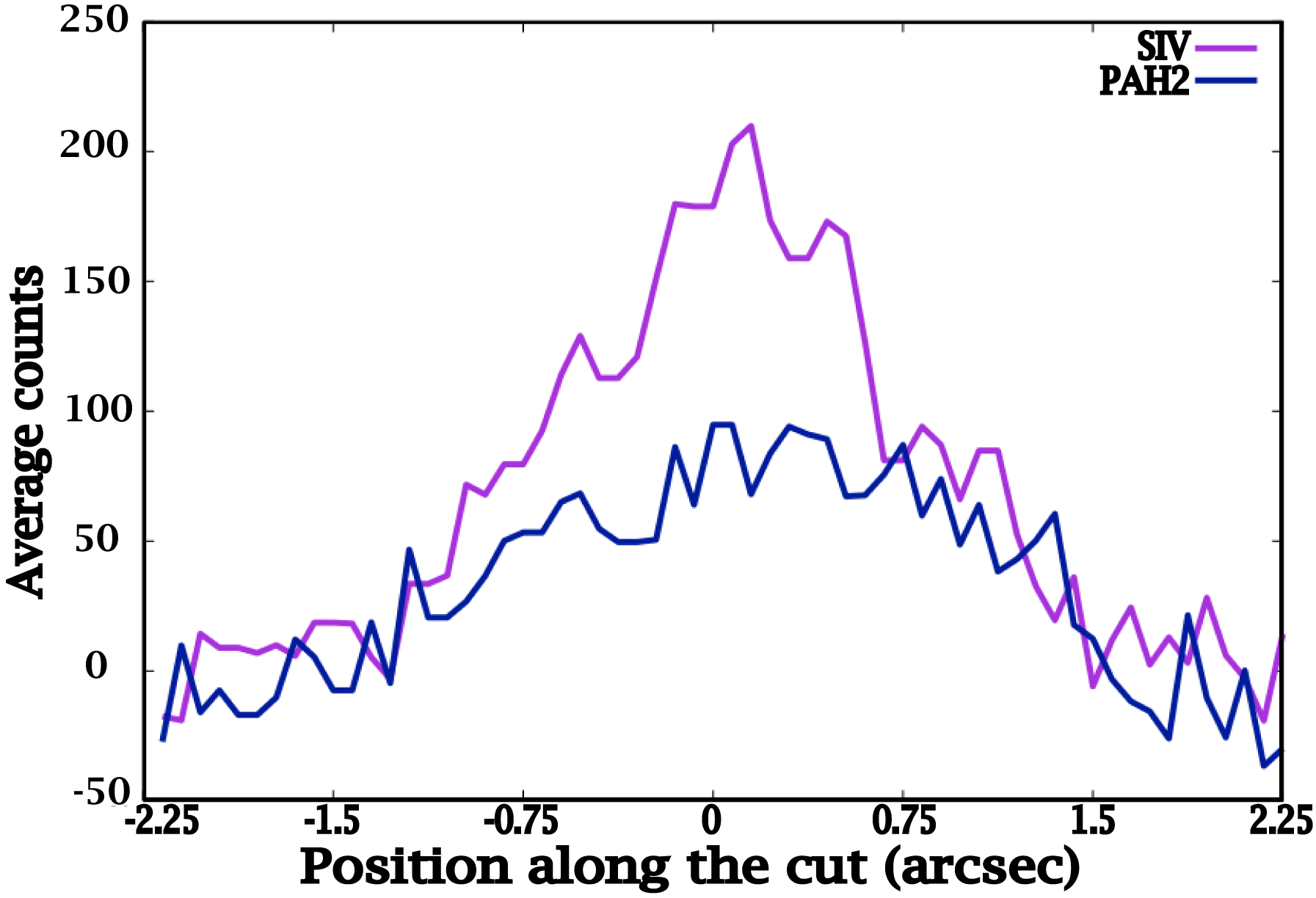}
     }
       \caption[VISIR image and spectra of M1-31]{PN M1-31: The leftmost image of the upper panel is an H$\alpha$ image taken with the {\it HST}. Next to it, we present the VISIR observations at 10.49 (SIV) and 11.25$\upmu$m (PAH2). North is up and East is left. The colour bar shows the intensity in terms of the rms. In the lower panel, the leftmost plot displays  a cut along the VISIR spatially resolved 2D spectra (PA=95$^{\circ}$), with emission from the 11.3$\upmu$m PAH band (green), the [SIV] emission line (red) and the continuum (black) shown. The next two plots are cuts made on the photometric images, in all the filters, at position angles corresponding to the torus (middle panel, cut made at PA=104$^{\circ}$) and outflows (right, cut made at PA=32$^{\circ}$).}
     \label{cuts}
\end{figure*}

{\bf M1-31} (PN G006.4+02.0, Fig. \ref{cuts})  The {\it HST} image shows an edge-on torus
 with outflows extending perpendicularly to this structure. In the VISIR images, we can see 
that the emission in the SIV filter follows the H$\alpha$ emission in the region occupied by 
the torus, with little emission in the outflows. In the PAH2 filter, the target is faint but
 we can see that the emission also traces the torus. Using the 2D resolved spectra, where the
 PA of the slit was 95$^{\circ}$ (approximately aligned with the torus), it appears that the
 emission of the PAHs comes from a broader region compared to the [SIV] line. This is also seen
 when making cuts along the VISIR images. For the plot in the middle we made a cut 
at a PA of 104$^{\circ}$ also along the torus. The SIV filter emission is brighter but less extended than the one 
in the PAH2 filter. On the rightmost plot we made a cut at a PA of 
32$^{\circ}$, along the outflows. For this position we can see a similar effect where the PAH2 
filter seems to have a broader emission, while the SIV filter are more concentrated towards the centre.

\begin{figure*}
 \centering
     \vbox{
      \hspace*{7.5cm}
\includegraphics[width=7.9cm, height=3.85cm]{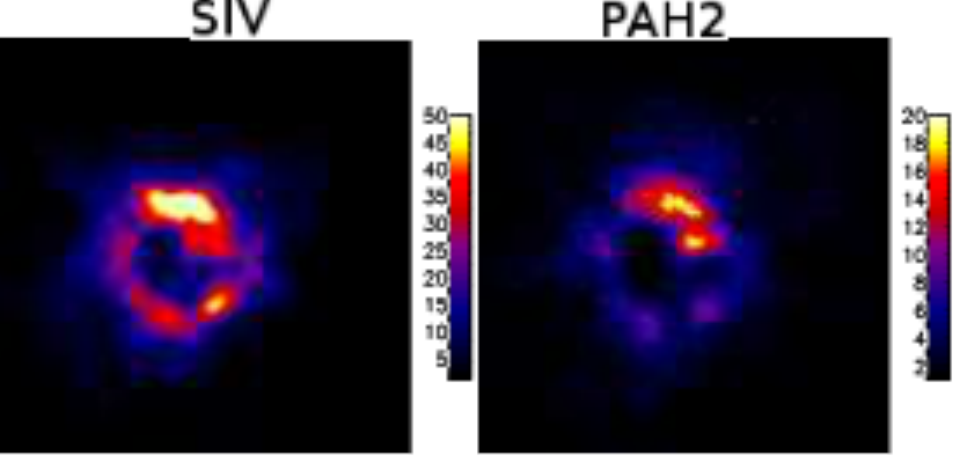}
               \includegraphics[width=5.7cm, height=3.85cm]{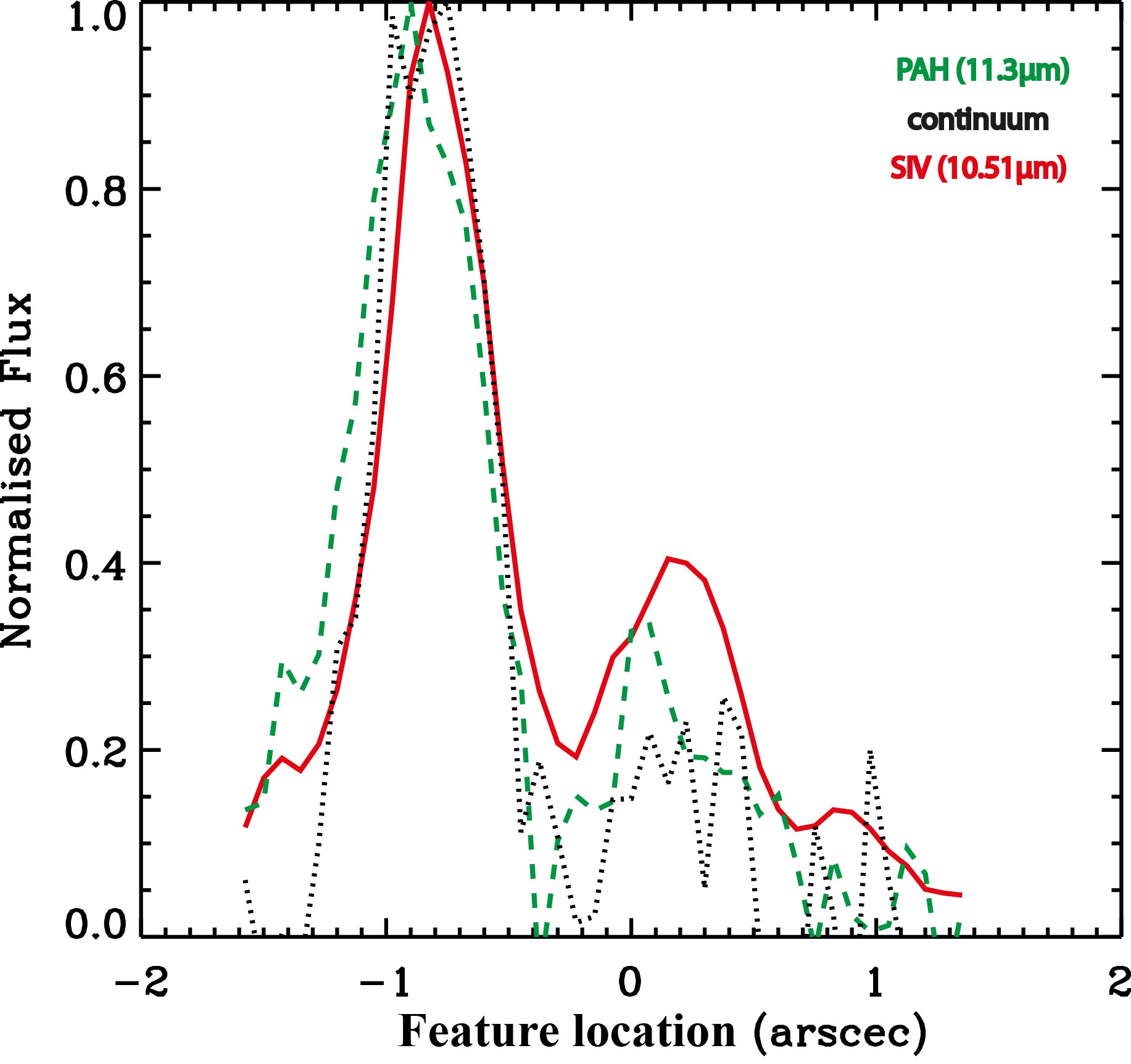}
               \includegraphics[width=5.7cm]{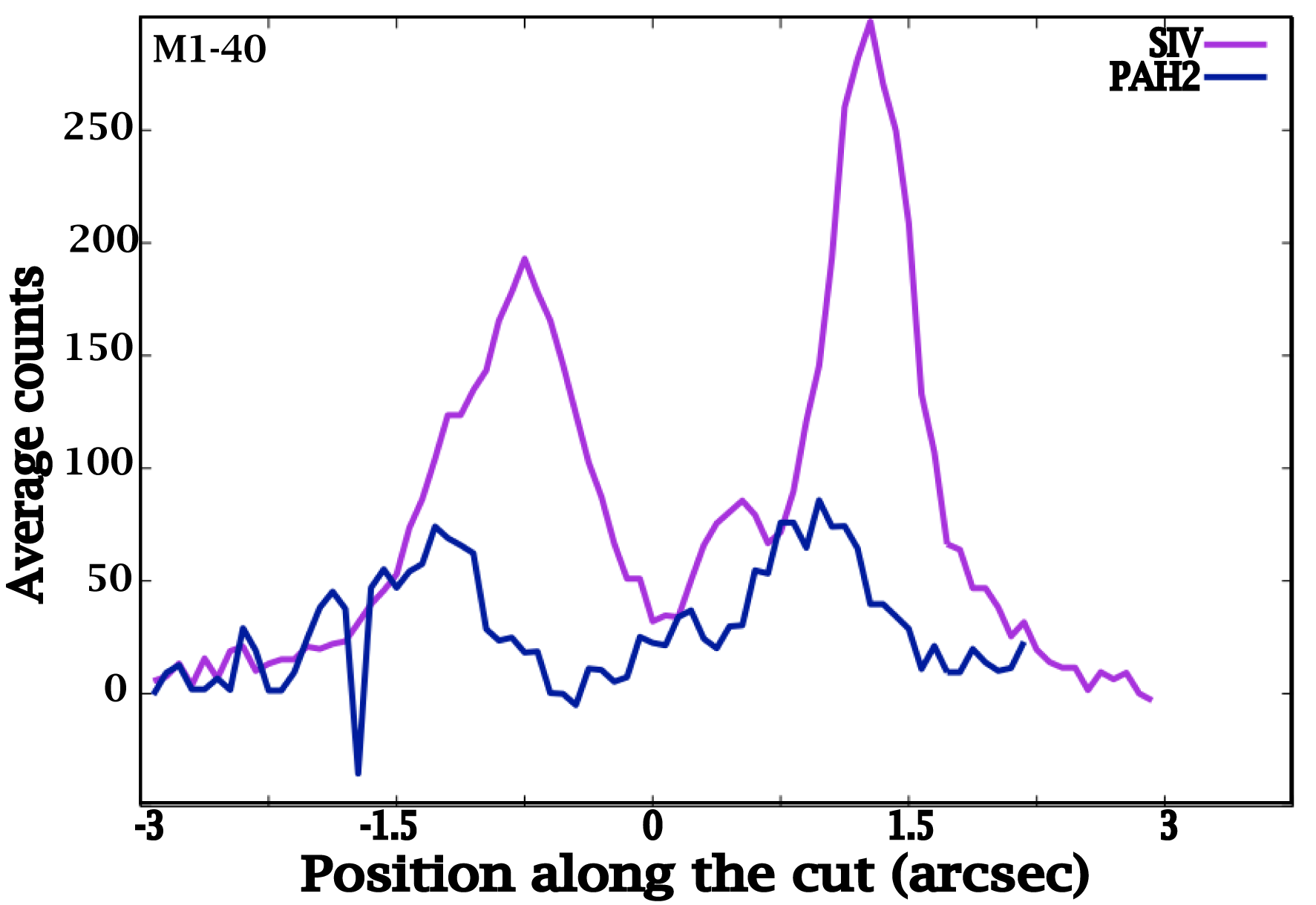}
		\includegraphics[width=5.7cm]{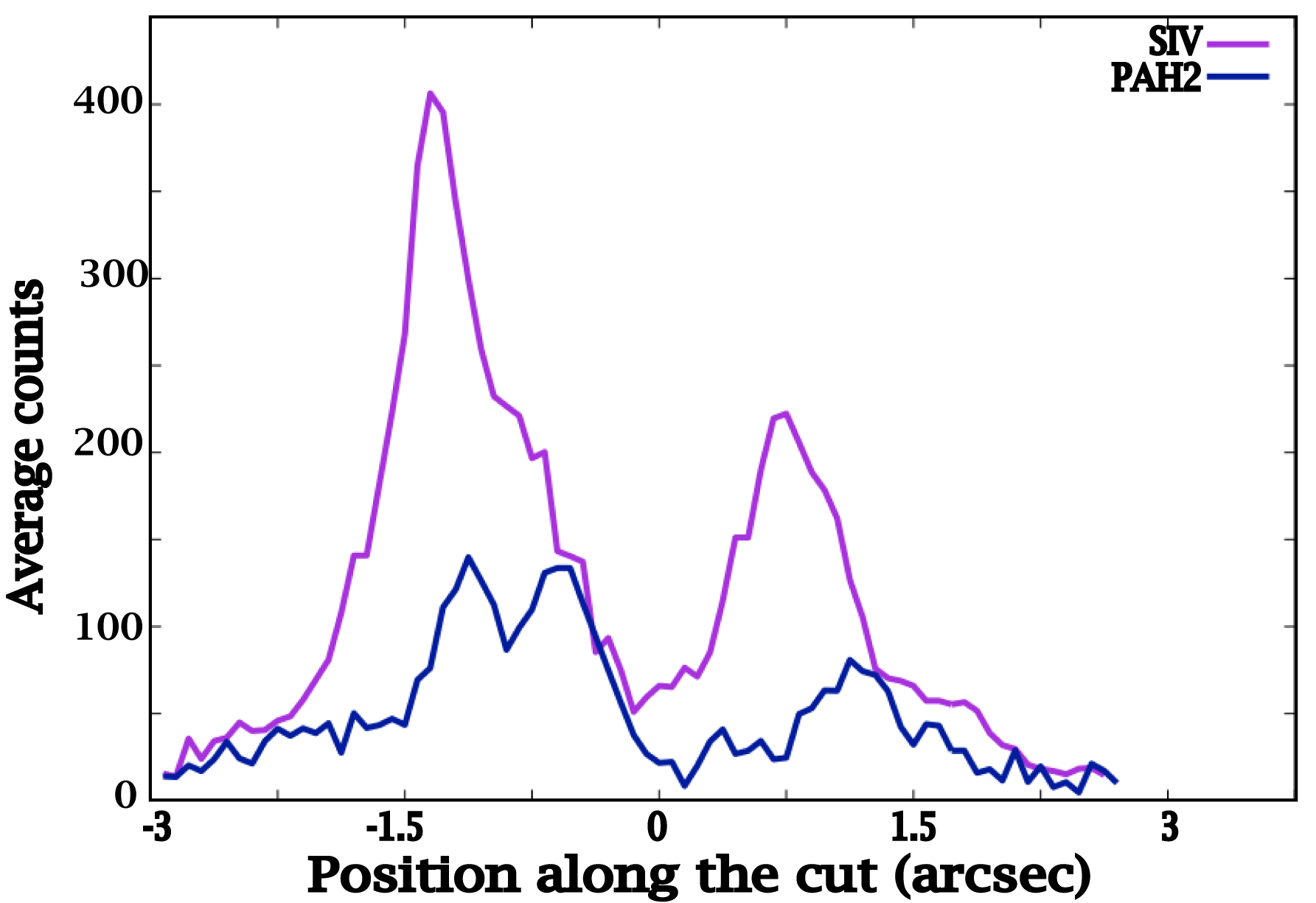}     
               }
                         \caption[VISIR image and spectra of M1-40]{PN M1-40: We present the VISIR observations at 10.49 (SIV) and 11.25$\upmu$m (PAH2). North is up and East is left. The colour bar shows the intensity in multiples of the rms. In the lower panel, the leftmost plot displays a cut along the VISIR spatially resolved 2D spectra (PA=-31$^{\circ}$), with emission from the 11.3$\upmu$m PAH band (green), the [SIV] emission line (red) and the continuum (black) shown. The next two plots are cuts made on the photometric images, in all the filters, at position angles corresponding to the torus (middle panel, cut made at PA=110$^{\circ}$) and outflows (right, cut made at PA=30$^{\circ}$).}
     \label{cuts_1}
\end{figure*}

{\bf M1-40} (PN G008.3-01.1, Fig. \ref{cuts_1}): The VISIR images reveal a clumpy doughnut-like structure seen close to face-on.
 In the SIV filter image, we can see the clumps and a  bright spot in the northern part of the nebula. In the
 PAH2 filter image, the clumps are visible too, although fainter, and the brightest emission  correlates in both filters.
 The spatial profile of a longslit spectrum, acquired with the slit aligned at a PA of -31$^\circ$,
 shows double-peaked emission (in [SIV], PAH emission and the continuum) originating from the bright northern and fainter southern
 rims of the nebular ring.  Cuts made along the photometric images at PAs of 110$^\circ$ and 30$^\circ$ show a very similar profile
 in all filters to that observed in the longslit spectrum. 
 
\begin{figure*}
       \centering
\vbox{
\includegraphics[width=15.5cm]{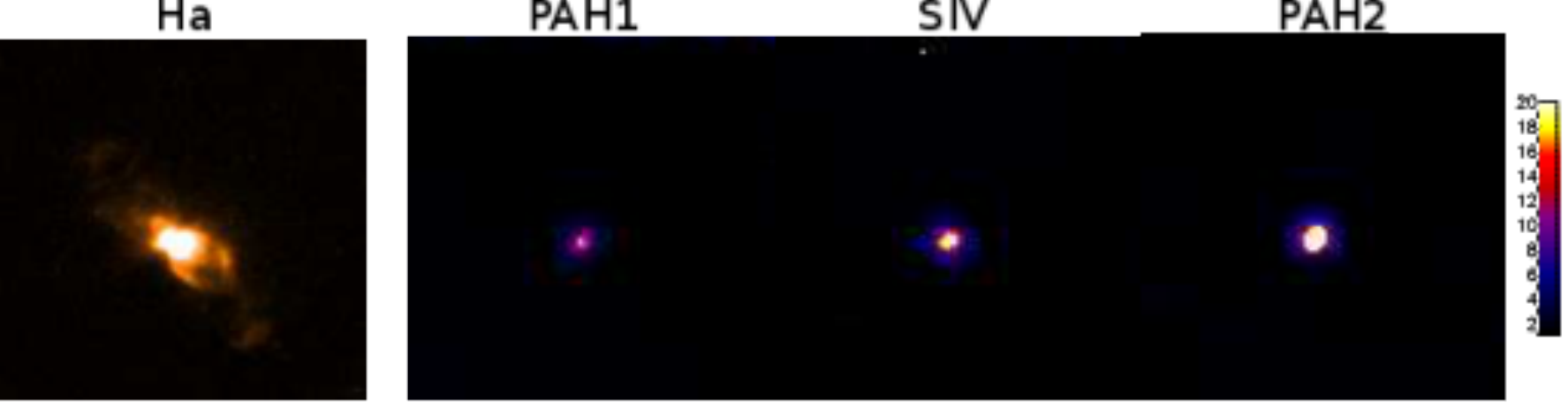}
     \includegraphics[width=5.7cm, height=3.85cm]{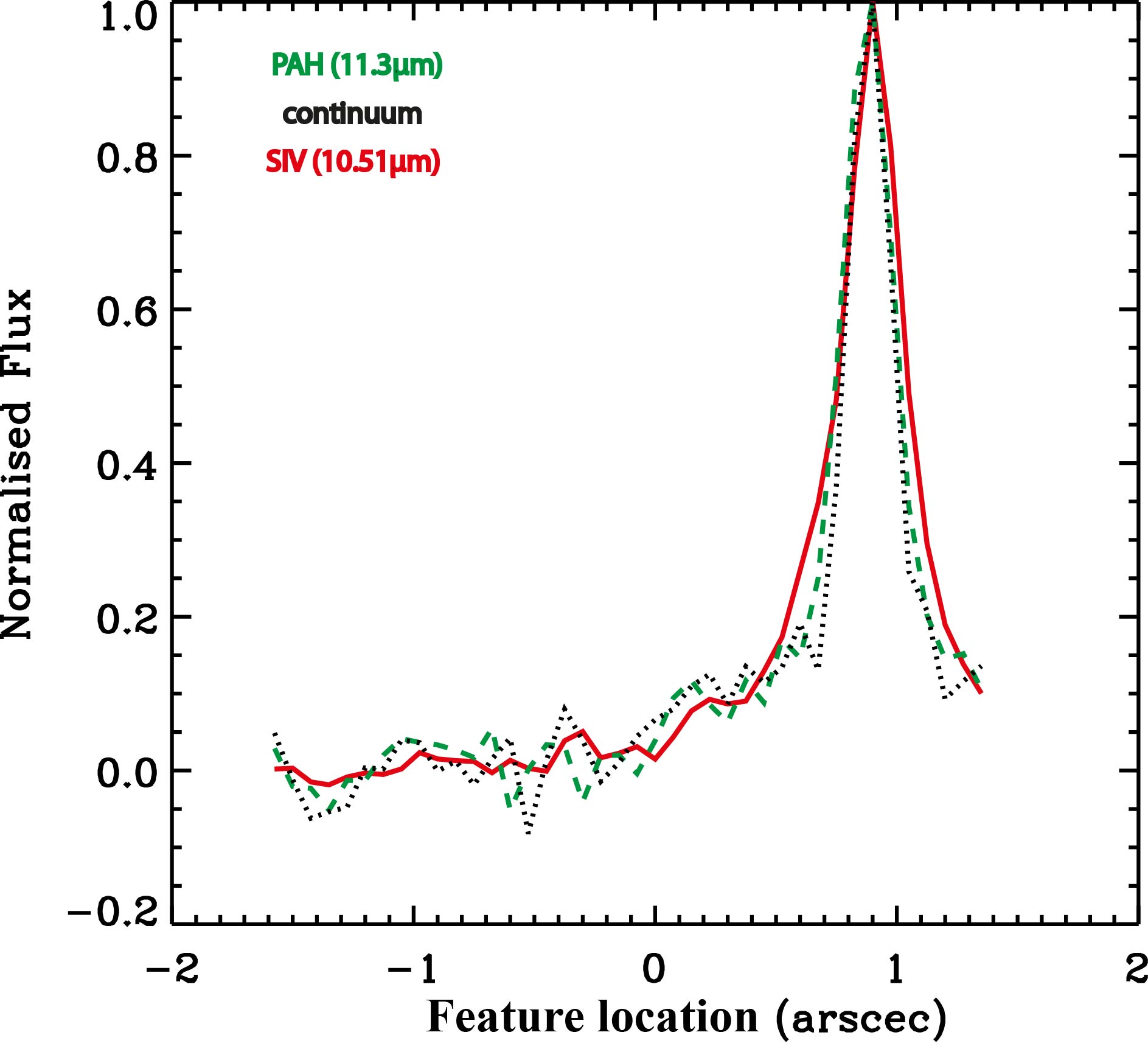}
          \includegraphics[width=5.7cm]{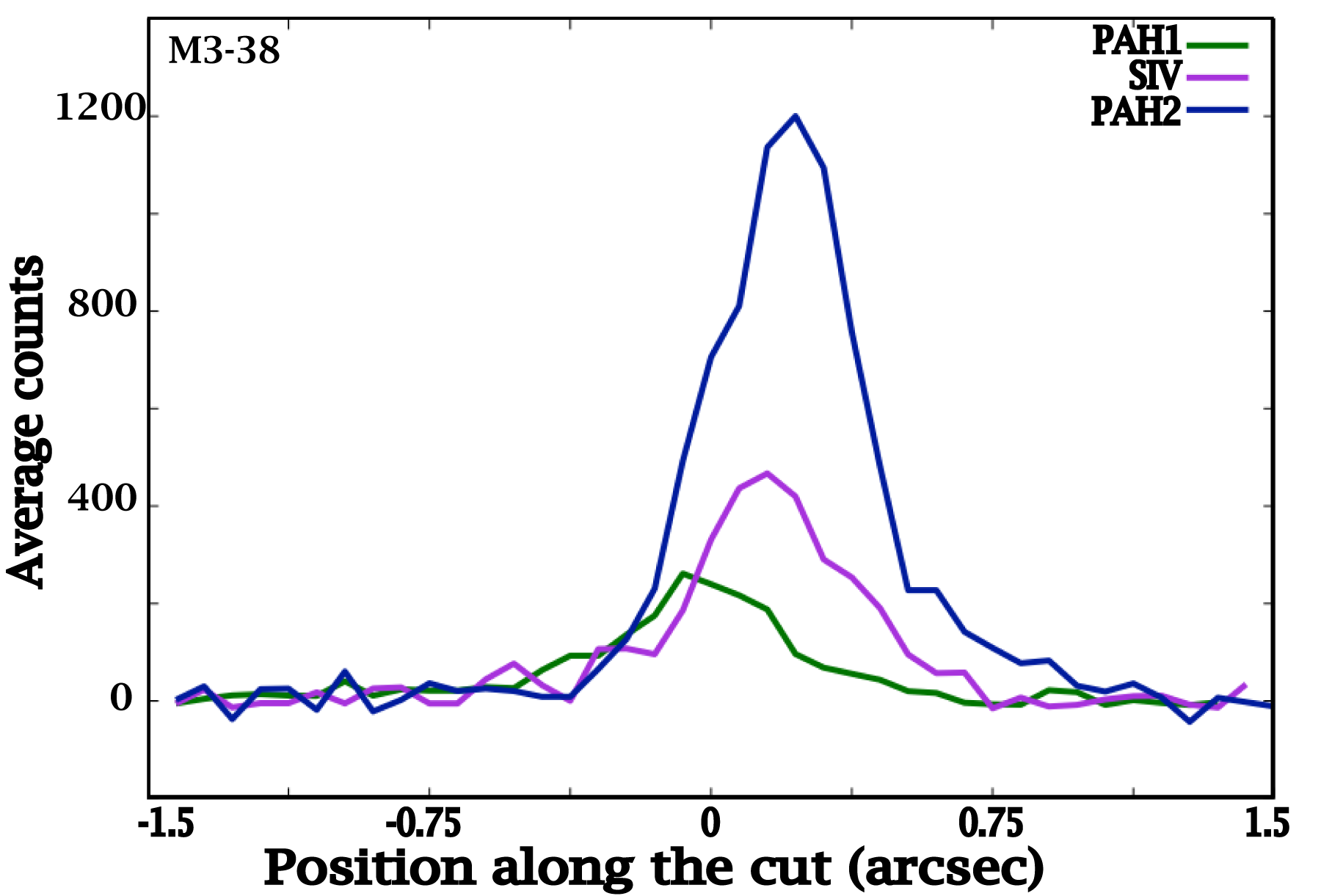}
	   \includegraphics[width=5.7cm]{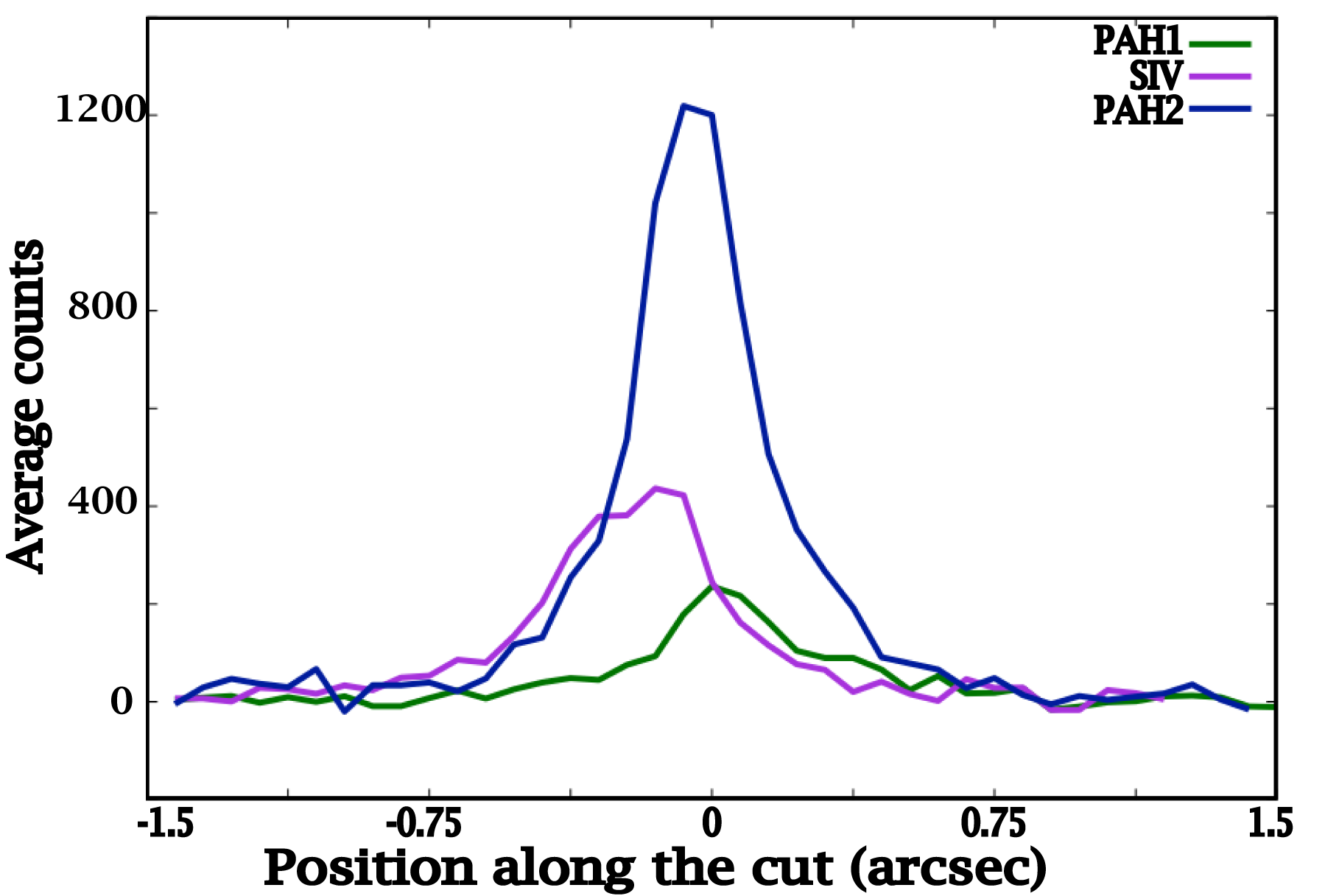}}
            \caption[VISIR image and spectra of M3-38]{PN M3-38: The leftmost image of the upper panel is an H$\alpha$ image taken with the {\it HST}. Next to it, we present the VISIR observations at 8.59 (PAH1),10.49 (SIV) and 11.25$\upmu$m (PAH2). North is up and East is left. The colour bar shows the intensity in multiples of the rms. In the lower panel, the leftmost plot displays a cut along the VISIR spatially resolved 2D spectra (PA=36$^{\circ}$), with emission from the 11.3$\upmu$m PAH band (green), the [SIV] emission line (red) and the continuum (black) shown. The next two plots are cuts made on the photometric images, in all the filters, at position angles corresponding to the torus (middle panel, cut made at PA=-20$^{\circ}$) and outflows (right, cut made at PA=70$^{\circ}$).}         
     \label{cuts_2}
\end{figure*}

{\bf M3-38} (PN G356.9+04.4) is a very compact PN. The H$\alpha$ {\it HST} image displayed Fig. \ref{cuts_2} shows the presence 
of a very bright central source (the torus), and  bipolar outflows almost seen almost edge-on. In the VISIR images we can not really 
resolve any structure, there is emission from the PN in all the three filters but it is very compact. In the bottom 
part of the Fig., we show three cuts. The left most plot is the cut of the spectral image along the spatial direction
 of the features (at a PA of 36$^{\circ}$), it shows that the PAHs seem to be in a slightly more compact region than the ionised region. On the
 contrary, the plot in the middle showing the cut on the photometric images at an PA of -20$^{\circ}$ (along the torus)
 presents the opposite scenario. The emission seems to be more extended in the PAH2 filter than in the SIV filter. Also the PAH1 filter emission's
 peak seems to be offset from the PAH2 and SIV filter peaks. In the leftmost plot, we present a cut along the outflows
 made at a PA of 70$^{\circ}$. This plot shows  that the emission  is more extended and brighter in the PAH2 filter than in the other
 filters, in this case the SIV filter shows the emission peak to be offset from the other two filters.

Using the spectral observations, we calculated the continuum contribution on each source by measuring 
the continuum level in the 3 segments of the N band.  For M1-31, the PAH feature have a 14$\sigma$ peak and the 
continuum contribution is on average 7\%. In the case of M1-40, the PAH feature were detected at a 27$\sigma$ and 
the continuum contribution represents on average 4\% of the signal. For M3-38 the continuum contribution is
 greater than for the other two sources and the detection level of the PAH feature is only at a 3$\sigma$, while 
the continuum for this PN contributes roughly 30\% of the emission.

\begin{figure*}
\centering
     \hbox{\includegraphics[width=8.2cm, height=7cm]{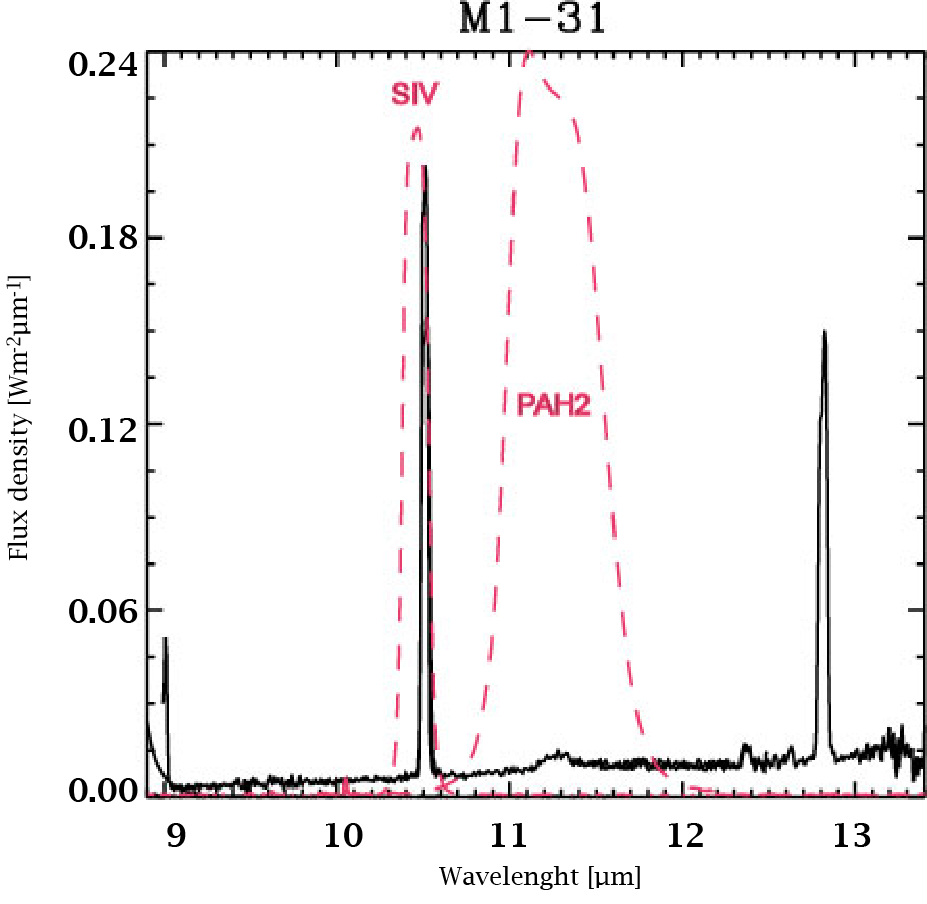}
              \hspace{1cm}
           \includegraphics[width=8.2cm, height=7cm]{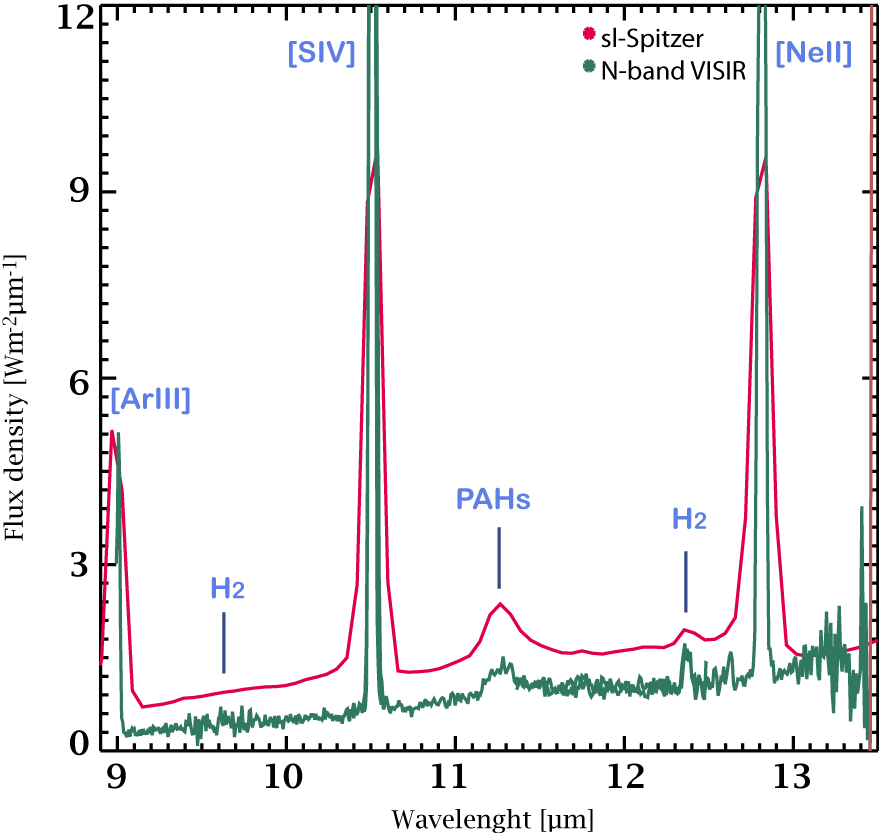}}
  \vspace{0.4cm}
     \hbox{\includegraphics[width=8.2cm, height=7.3cm]{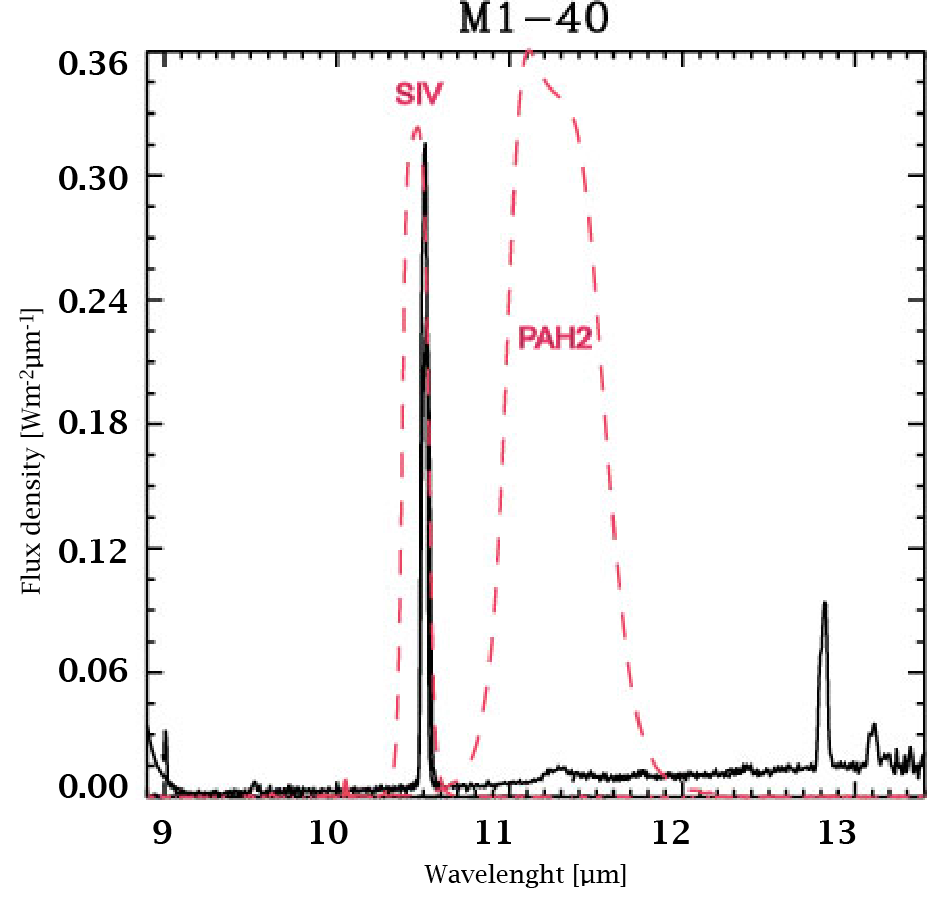}  
          \hspace{1cm}
        \includegraphics[width=8.2cm, height=7cm]{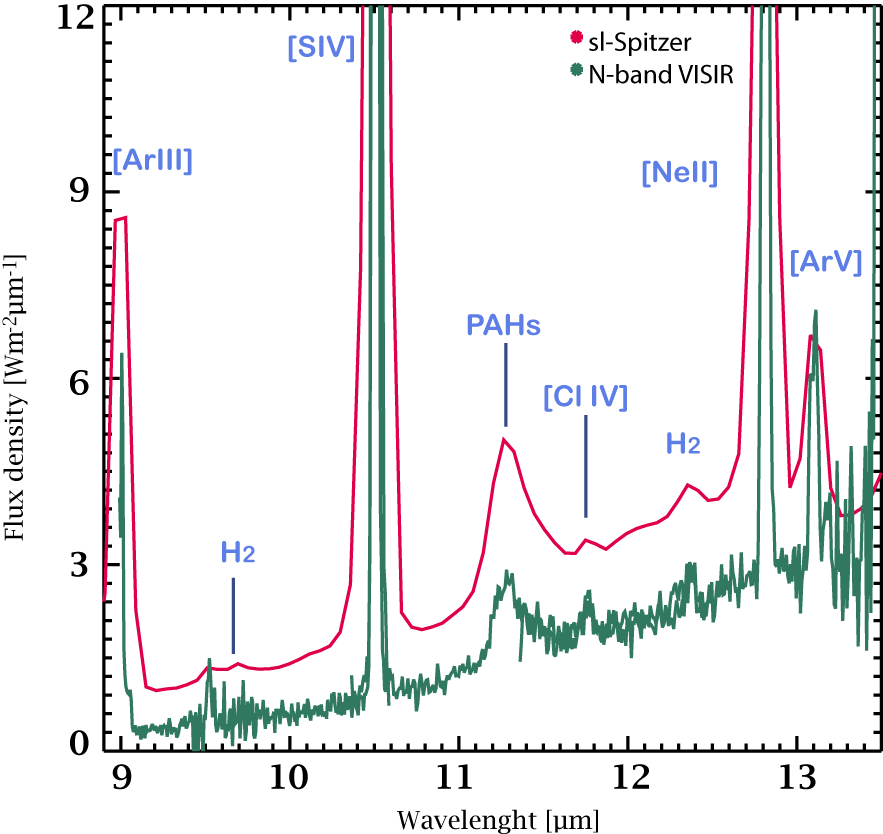}}
  \vspace{0.4cm}
     \hbox{\includegraphics[width=8.2cm, height=7.3cm]{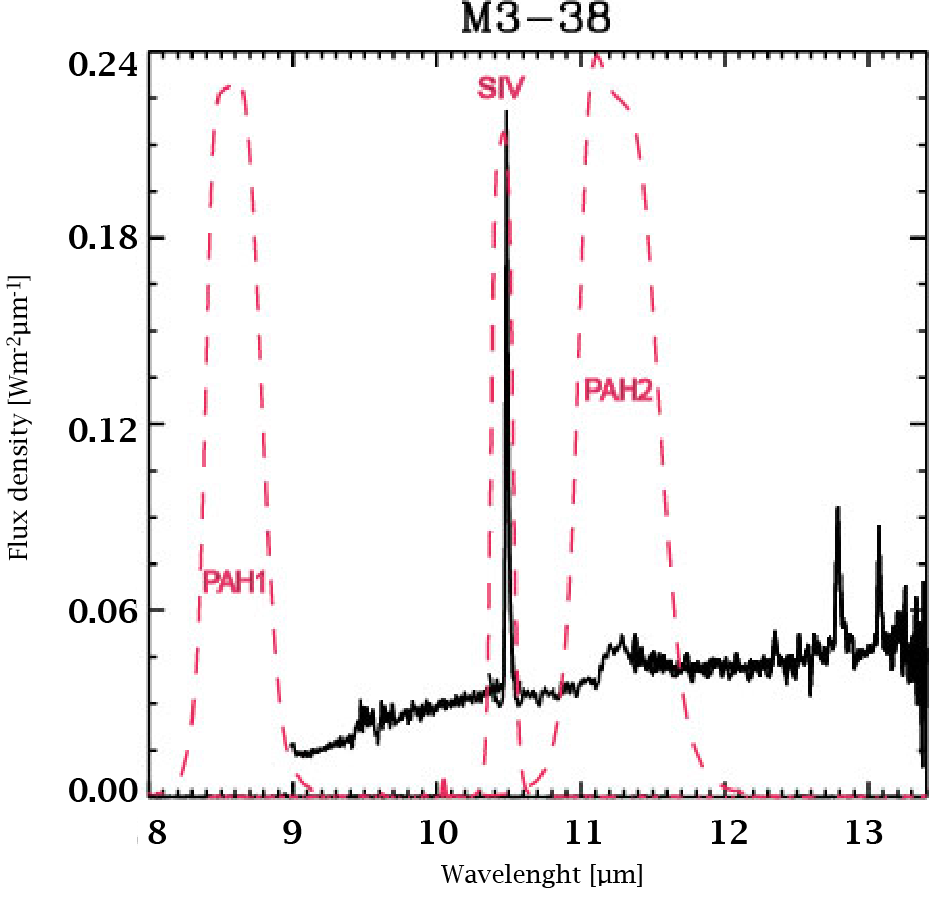}
          \hspace{1cm}
          \includegraphics[width=8.2cm, height=7cm]{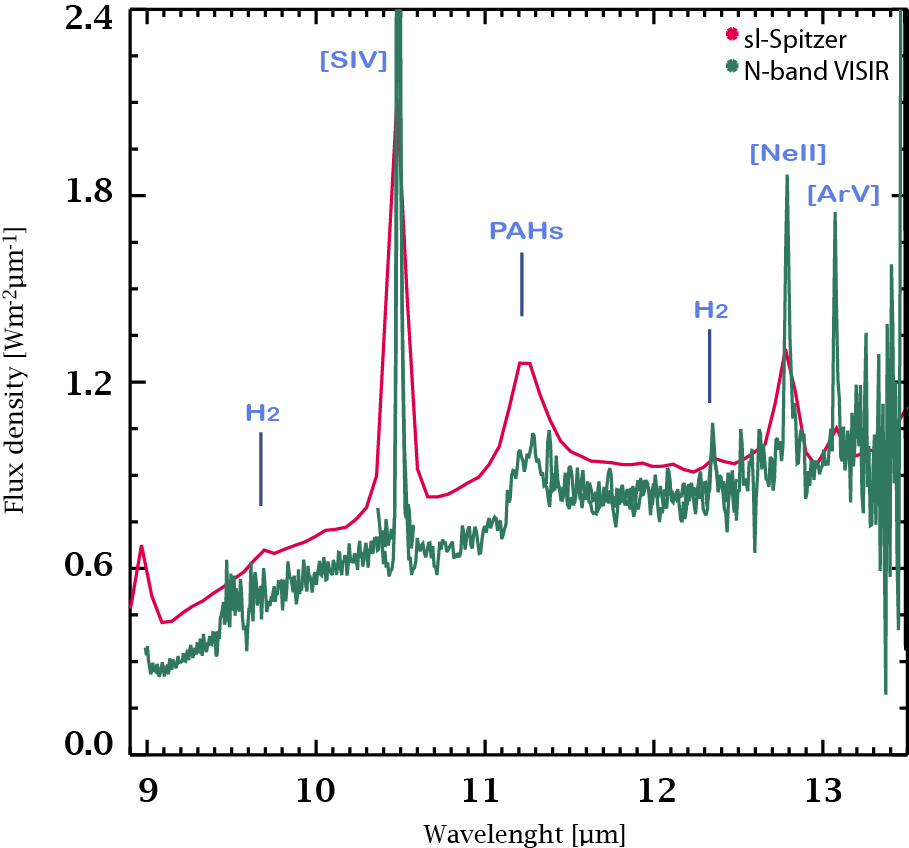}}
               \caption[VISIR and {\it Spitzer} spectra]{Left panel shows the {\it VLT}-VISIR N-band spectrum of the PNe M1-31, M1-40 and M3-38 with the absolute transmission curves for the VISIR PAH1, SIV, and PAH2 filters over-plotted (dashed pink line). The right panel shows the {\it VLT} and {\it Spitzer} spectra over-plotted, with emission line and PAH bands identified. The VISIR spectra were multiplied by a factor of 100 for M1-31 and M1-40, and a factor of 50 for M3-38 so in order to be be compared with the {\it Spitzer} spectra.}
     \label{filters}
\end{figure*}

In Fig. \ref{filters}, we present the VISIR spectra of the PNe M1-31, M1-40, 
and M3-38 with the absolute filter transmission curves for the PAH1, SIV and
 PAH2 filters over-plotted, as well as comparing the VISIR spectra with those 
acquired using {\it Spitzer}.  The spectrum of M1-31 shows the 11.3$\upmu$m
 feature attributed to PAHs, and the H$_2$ line at 9.6 and 12.4$\upmu$m.
The spectrum for the PN M1-40 displays [ArV] emission, indicating that this is a high-excitation
 PN, while also showing PAH feature at 11.3, the [Cl IV] line at 11.74$\upmu$m, and H$_2$ at 9.6 and 12.4$\upmu$m. 
M3-38 is also a high-excitation PN, as its spectrum also shows the [ArV] emission line. The 
spectrum displays an excess in the continuum due to the very broad silicate feature at 10$\upmu$m. H$_2$ emission is also detected at 9.6 and 12.4$\upmu$m.

\subsection{Emission lines \& PAHs}

Using spectra taken with {\it Spitzer} from \cite{me11}, we measured the peak of the [SIV] emission line, 
centred at 10.51$\upmu$m, [NeII] centred at 12.81$\upmu$m and [NeIII] centred at 15.6$\upmu$m, as well as the  
integrated flux of the PAH feature at 11.3$\upmu$m and the H$_2$ line at 12.4$\upmu$m. 
The uncertainty on the measurements of the flux density are less than 1\% due to the very high S/N ratio obtained with {\it Spitzer}.
Table \ref{intense} lists the strength of  these lines and the PAHs
 features, the measured size at full-width of half-maximum of each PN (in arcseconds), using the SIV filter of the VISIR images, the uncertainty on the 
size measurements is less than 5\%. The temperature 
(in K) is also presented, where the uncertainty on the temperature measurements is our biggest source of
 uncertainty. The temperature values are an approximation of the effective temperature of the central star, 
and are the black-body temperature derived from photoionisation modelling of the nebulae by \citet{gesicki06}. The estimated accuracy in 
the determination of the temperature is 10\% (see \citealp{gesicki06} for more details).

\begin{table*}
\caption{Integrated flux for the 11.3$\mu$m PAH feature and the H$_2$ line
 (12.4$\upmu$m), and emission lines ratios for the [SIV], [NeII], and [NeIII] lines. Size and the effective temperature of the central star is also presented. }
\label{intense}
\centering
\begin{tabular}{llcccccc} 
\hline\hline Name & PNG Name &  11.3$\upmu$m  & H$_2$ (12.4$\upmu$m) & Size & log([SIV]/[NeII]) & log([NeIII]/[NeII]) &  log(T$_{eff}$)\\
 & & 10$^{-12}$Wm$^{-2}$ & 10$^{-12}$Wm$^{-2}$ & arcsec & & & K \\
 \hline
Cn1-5 & 002.2 $-$09.4 & 5.28$\pm$0.052 & 0.45$\pm$0.004 & 5$\pm$0.25 & $-$0.07$\pm$0.001 & 0.84$\pm$0.011 & 5.04$\pm$0.04$^1$ \\	
M1-25 & 004.9 $+$04.9 & 2.07$\pm$0.020 & 0.27$\pm$0.003 & 3$\pm$0.15 & $-$1.39$\pm$0.019 & 0.87$\pm$0.012 & 4.62$\pm$0.04$^2$ \\
M1-31 & 006.4 $+$02.0 & 3.99$\pm$0.039 & 0.36$\pm$0.003 & 1.6$\pm$0.08 & 0.14$\pm$0.002 & 0.98$\pm$0.013 & 4.68$\pm$0.04$^3$ \\
H1-61 & 006.5 $-$03.1 & 4.44$\pm$0.044 & 0.21$\pm$0.002 & 1.4$\pm$0.07 & $-$1.22$\pm$0.017 & 0.11$\pm$0.001 & 4.70$\pm$0.04$^4$ \\
M3-15 & 006.8 $+$04.1 & 0.66$\pm$0.006 & 0.27$\pm$0.002 & 3$\pm$0.15 & 0.66$\pm$0.009 & 1.35$\pm$0.018 & 4.92$\pm$0.04$^3$ \\
Hb6 & 007.2 $+$01.8 & 1.50$\pm$0.015 & 0.99$\pm$0.009 & 5$\pm$0.25 & 1.03$\pm$0.014 & 1.45$\pm$0.021 & 4.84$\pm$0.04$^5$ \\
M1-40 & 008.3 $-$01.1 & 10.89$\pm$0.108 & 0.84$\pm$0.008 & 3.5$\pm$0.17 & 0.35$\pm$0.005 & 0.82$\pm$0.011 & 5.15$\pm$0.04$^2$ \\
Th3-4 & 354.5 $+$03.3 & 1.56$\pm$0.015 & 0.42$\pm$0.004 & 1.2$\pm$0.06 & 0.63$\pm$0.008 & 1.45$\pm$0.021 & 4.68$\pm$0.04$^3$ \\
M3-38 & 356.9 $+$04.4 & 1.92$\pm$0.029 & 0.06$\pm$0.001 & 1.1$\pm$0.05 & 0.45$\pm$0.006 & 1$\pm$0.014 & 5.09$\pm$0.04$^3$ \\
H1-43 & 357.1 $-$04.7 & 2.22$\pm$0.022 & 0.09$\pm$0.001 & 1.1$\pm$0.05 & $-$1.61$\pm$0.022 & -- \,  \,  -- & 4.26$\pm$0.04$^3$\\
H1-40 & 359.7 $-$02.6 & 2.07$\pm$0.020 & 0.27$\pm$0.002 & 1.2$\pm$0.06 & 0.12$\pm$0.001 & 0.84$\pm$0.011 & 4.85$\pm$0.04$^6$ \\
 \hline
  \end{tabular} \\
$^1$\cite{acker03}, $^2$\cite{acker02}, $^3$(Gesicki et al 2013 subm.), \\ $^4$(Gesicki priv.comm. from simple
photoioization model) $^5$\cite{phillips03}, $^6$\cite{gesicki00}
\end{table*}

Trying to find a diagnostic of the ionisation state of emission-line objects,
 \citet{groves08}  found a strong correlation between the [SIV]/[NeII] versus 
the [NeIII]/[NeII] emission line ratios. The authors were looking for a 
ground-based technique to analyse the ionisation state of a variety of 
astronomical objects. They  found a correlation with a linear fit of 
log([NeIII]/[NeII]) = 0.81log([SIV]/[NeII])+0.36 accurate to a 
factor of two over four orders of magnitude. This correlation was 
observed for Galactic and extragalactic HII regions, ultra luminous 
infra-red galaxies (ULIRGs), active galactic nuclei (AGN), Galactic and extragalactic planetary nebulae (PNe), starburst galaxies, and blue compact dwarfs (BCDs). 

\begin{figure*}
\centering
\includegraphics[width=14cm, height=8cm]{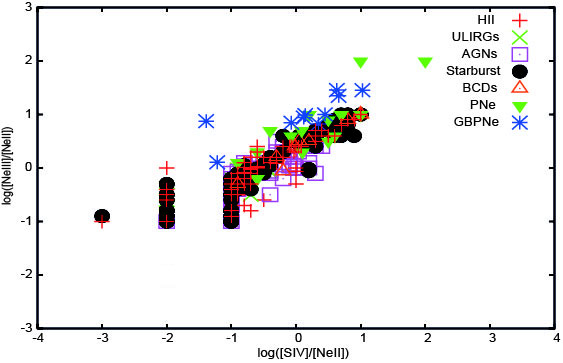}
\caption[Line ratios correlation]{[SIV]/[NeII] - [NeIII]/[NeII] correlation for Galactic Bulge Planetary Nebulae (GBPNe).}
     \label{lineRatios}
\end{figure*}

For the 11 objects presented in this work, line intensities were 
measured and  plotted in the same manner to that in \citet{groves08},
 with the relationship presented in Fig. \ref{lineRatios}. To compare the GBPNe
 line intensities with all the other emission-line objects, we used data 
 kindly provided by  the first author of the above mentioned paper
 (see also references therein). We find that the [SIV]/[NeII] - [NeIII]/[NeII] correlation also works 
for PNe in the Galactic Bulge (as predicted by photoionisation theory, higher ionisation leads to higher
 values for both ratios). It is interesting to mention that even though GBPNe follow the same correlation,
 they all seem to reside in the upper part of the plot, in fact, this is true also for Galactic PNe, 
although they present the same slope, most of them seem to have higher values of the [NeIII]/[NeII] ratio.

\begin{figure*}
   \hbox{\includegraphics[width=8.4cm]{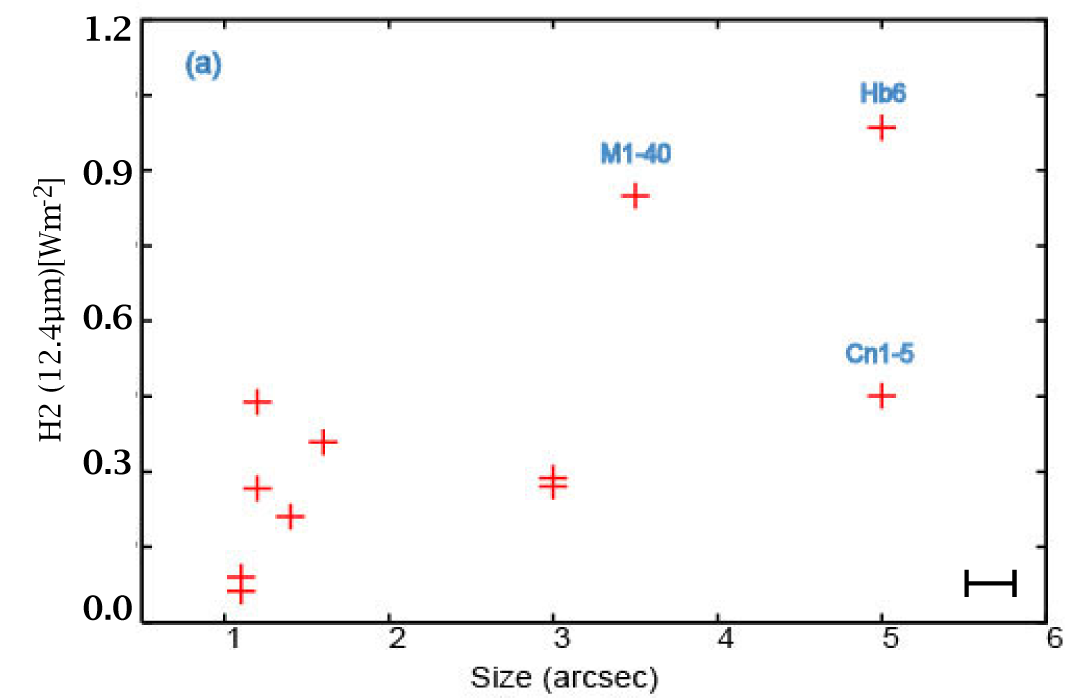}
         \hspace{0.01cm}
     \includegraphics[width=8.5cm]{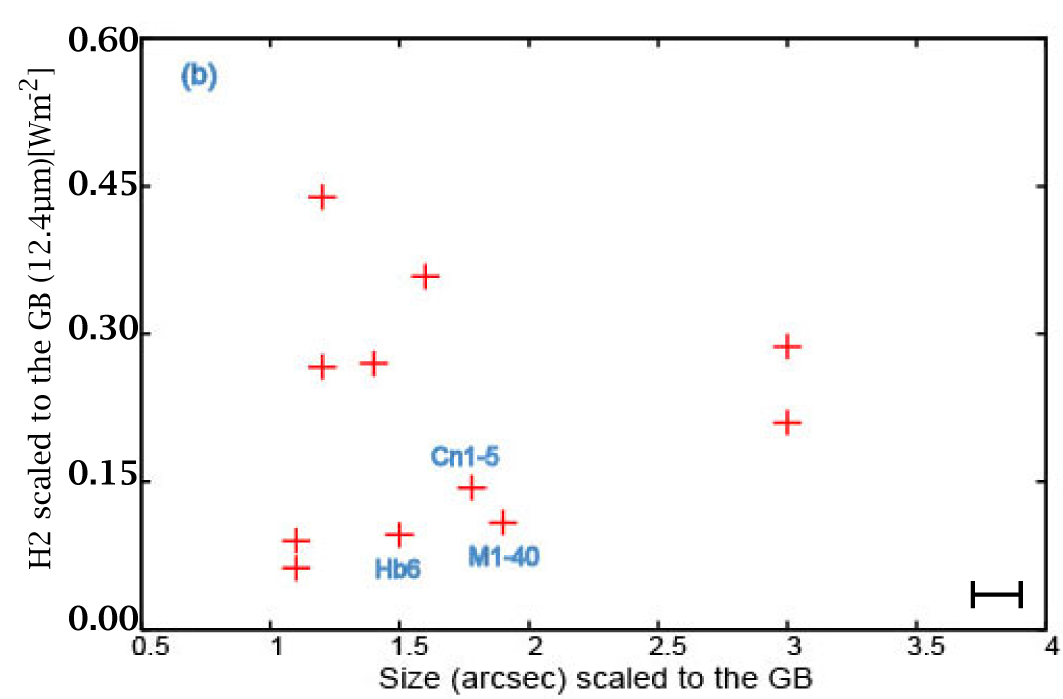}}
\vspace{0.7cm}
   \hbox{\includegraphics[width=8.5cm]{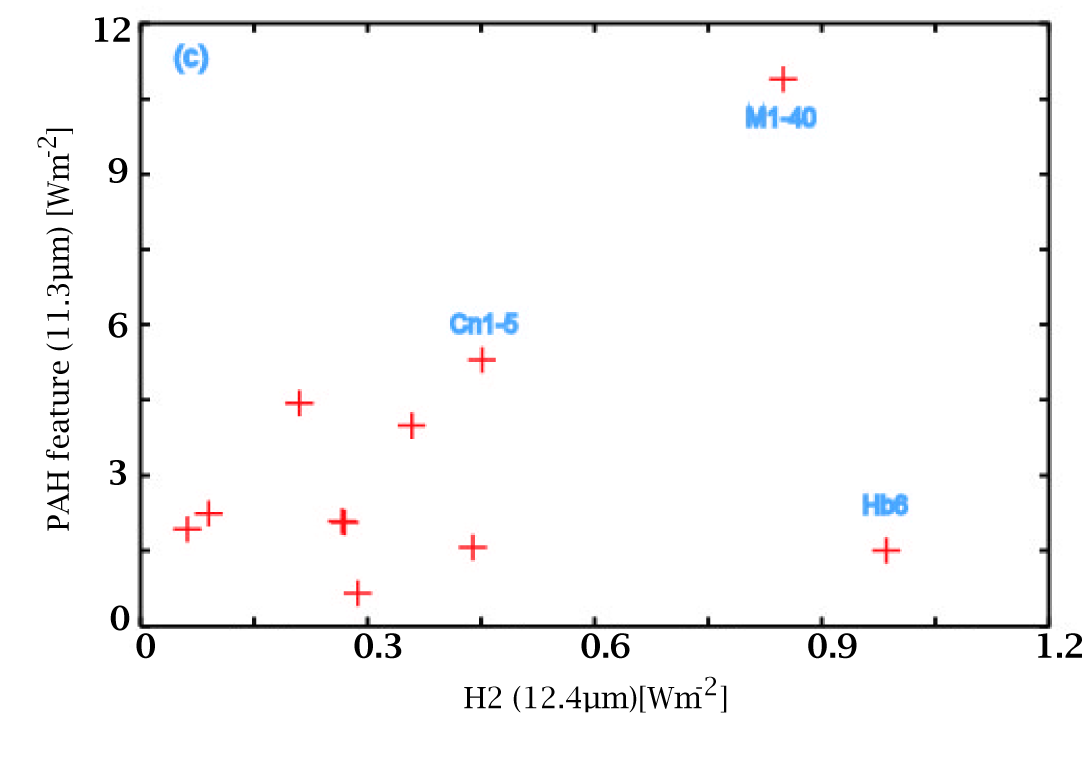}
         \hspace{0.01cm}
     \includegraphics[width=8.5cm]{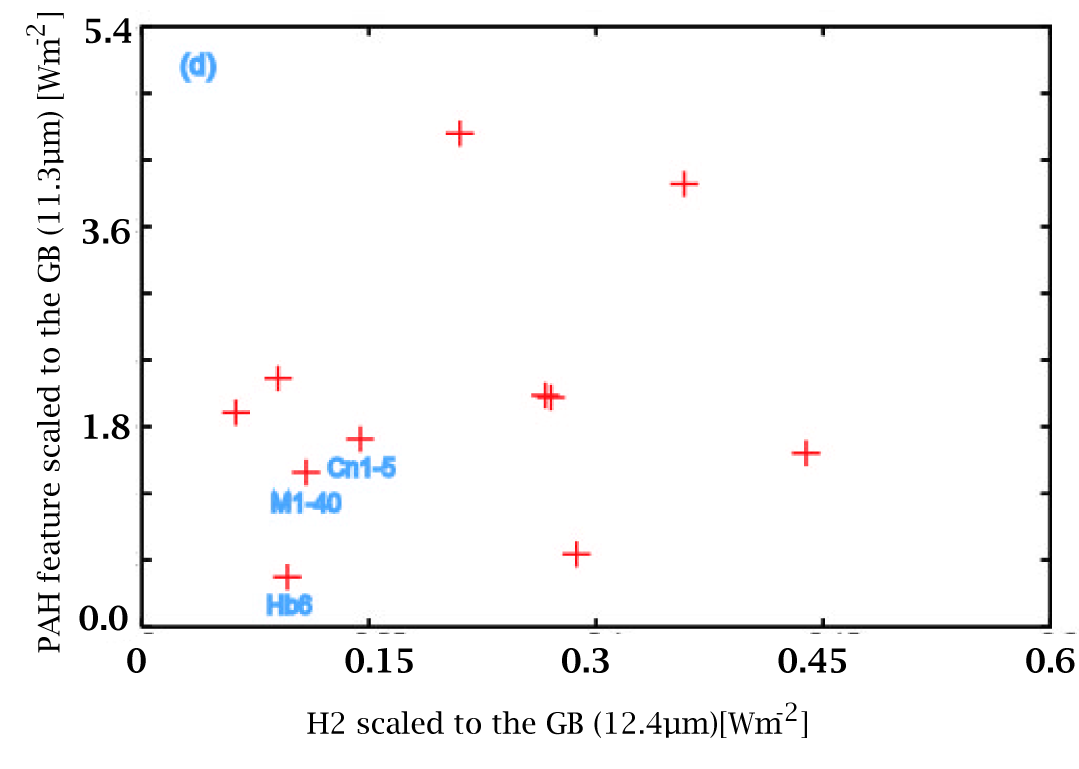}}
\vspace{0.7cm}
   \hbox{
     \hspace{0.05cm}
   \includegraphics[width=8.2cm]{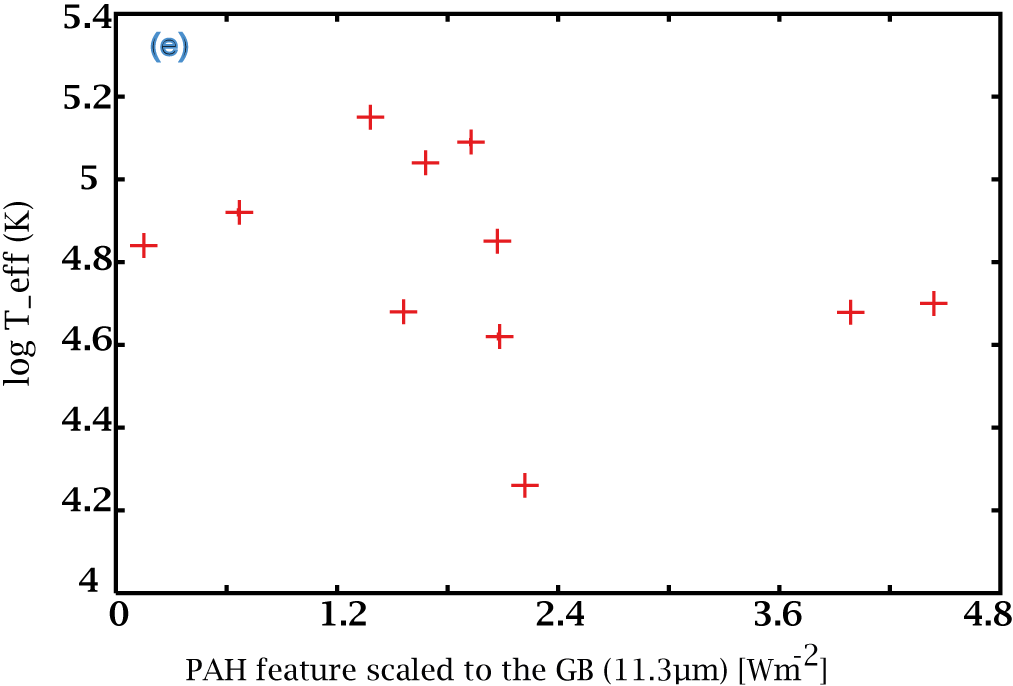}
        \hspace{0.05cm}
     \includegraphics[width=8.4cm, height=5.5cm]{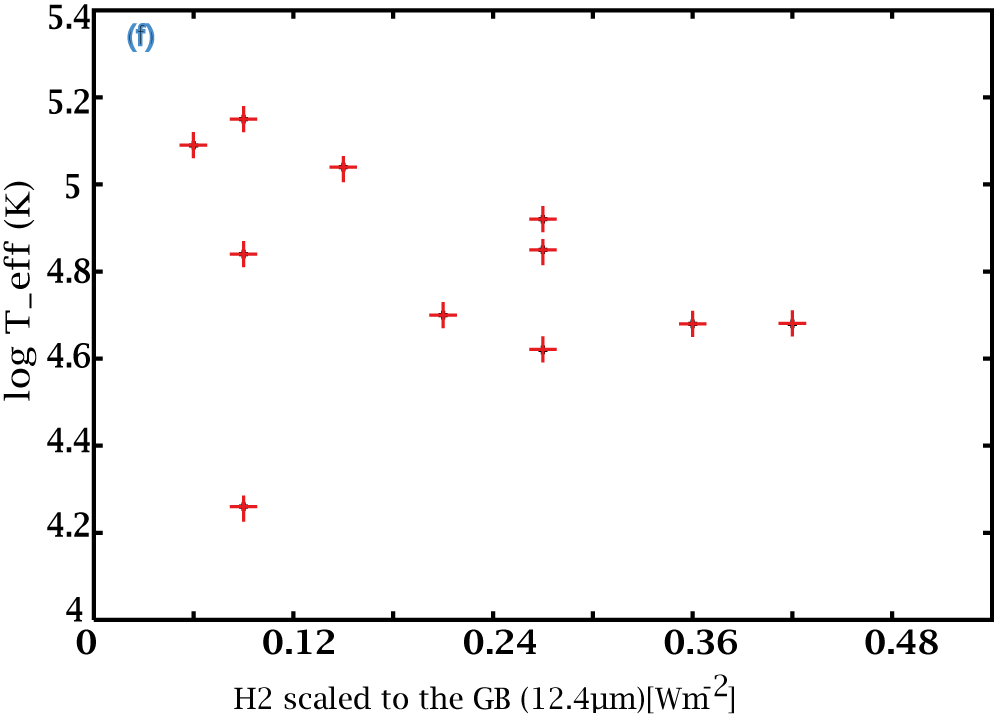}}
      \caption[Size vs H$_2$]{In panel (a) we show the size of the PNe in arcseconds versus its  H$_2$ emission line strength (12.4$\upmu$m). We clearly see that the three objects non-GBPNe lie away from the rest of the sample, the bar on the right-bottom corner represents the error bar for the size measurements, for the fluxes, the error bar is smaller than the symbol. In panel (b), we correct the distances of these non-GBPNe to the distances of the bulge, showing in general PNe from the GB have stronger H$_2$ (12.4$\upmu$m) emission compared to those in the disk, the bar on the right-bottom corner represents the error bar for the size measurements, as for the fluxes, the error bar is smaller than the symbol. \\
Panel (c) shows the H$_2$ emission lines strength at 12.4$\upmu$m compared with the 11.3$\upmu$m PAH band. Panel (d) shows the same emission features but scaling the  values for the PNe Hb6, Cn1-5 and M1-40 to the distance of the GB (8.5 kpc).\\
Panel (e) shows the integrated flux of the 11.3$\upmu$m PAH band versus the effective temperature of the central star, indicating no correlation. For panel (f) we present the strength of the H$_2$ emission line at 12.4$\upmu$m versus the effective temperature of the central star, displaying a slight anti-correlation is seen, with H$_2$ flux decreasing with increasing central star temperature.}
   \label{12Temp}
\end{figure*}

Using the values in Table \ref{intense}, we compared the PAH integrated flux (at 11.3$\upmu$m) 
and H$_2$ (at 12.4$\upmu$m) with the PN size in arcseconds, the central star
 effective temperature, and the [SIV], [NeII], [NeIII] emission line intensity, with their corresponding errors. Fig. \ref{12Temp} (a) displays a 
plot of the PN size versus H$_2$ (12.4$\upmu$m) emission line intensity. We clearly see the three objects that are outliers, the bar in the right-bottom corner represents the error in the size measurements. 
Despite a certain level of scatter,  the PNe Hb6, M1-40
 and Cn1-5 are far from the region occupied by the other PNe studied. This leads us to 
conclude that they are at a different distance, and therefore not in the GB - consistent 
with the results of \cite{me11}, who find that these PNe are found at distances much closer 
than the GB. According to \citet{phillips11}, Hb6 is at a distance of 2.5 kpc, Cn1-5 at 4.5
 kpc and M1-40 at 2.8 kpc.  We, therefore, scaled their sizes and surface brightness to the
 distance of the Bulge (8.5 kpc), in order to consistently compare their intensities and sizes
 to the other PNe in our sample. Table \ref{intense2} shows the values presented in 
table \ref{intense} re-scaled to place PNe Hb6, Cn1-5 and M1-40 in the GB, while a
 replotted relationship between the strength of the H$_2$ line (at 12.4$\upmu$m) and 
the PN size is shown in Fig. \ref{12Temp} panel (b),  we can see that, in general, PNe from the GB have stronger 12.4$\upmu$m H$_2$ emission lines compared to the non-GB ones.\\

\begin{table*}
\caption{Same as Table \ref{intense} but with Hb6, Cn1-5 and M1-40 fluxes and sizes scaled to the distance of the Galactic Bulge (8.5 kpc). }
\label{intense2}
\centering
\begin{tabular}{llcccccc} 
\hline\hline Name & PNG Name &  11.3$\upmu$m  &  H$_2$ (12.4$\upmu$m) & Size & log([SIV]/[NeII]) & log([NeIII]/[NeII]) &  log($T_{\rm{eff}}$)\\
 & & 10$^{-12}$Wm$^{-2}$ & 10$^{-12}$Wm$^{-2}$  & arcsec & & & K \\
 \hline
Cn1-5 & 002.2 $-$09.4 & 1.68$\pm$0.016 & 0.15$\pm$0.001 & 1.78$\pm$0.09 & $-$0.07$\pm$0.001 & 0.84$\pm$0.011 & 5.04$\pm$0.04$^1$ \\	
M1-25 & 004.9 $+$04.9 & 2.07$\pm$0.020 & 0.27$\pm$0.003 & 3$\pm$0.15 & $-$1.39$\pm$0.019 & 0.87$\pm$0.012 & 4.62$\pm$0.04$^2$ \\
M1-31 & 006.4 $+$02.0 & 3.99$\pm$0.039 & 0.36$\pm$0.003 & 1.6$\pm$0.08 & 0.14$\pm$0.002 & 0.98$\pm$0.013 & 4.68$\pm$0.04$^3$ \\
H1-61 & 006.5 $-$03.1 & 4.44$\pm$0.044 & 0.21$\pm$0.002 & 1.4$\pm$0.07 & $-$1.22$\pm$0.017 & 0.11$\pm$0.001 & 4.70$\pm$0.04$^4$ \\
M3-15 & 006.8 $+$04.1 & 0.66$\pm$0.006 & 0.27$\pm$0.002 & 3$\pm$0.15 & 0.66$\pm$0.009 & 1.35$\pm$0.018 & 4.92$\pm$0.04$^3$ \\
Hb6 & 007.2 $+$01.8 & 0.15$\pm$0.001 & 0.09$\pm$0.001 & 1.5$\pm$0.75 & 1.03$\pm$0.014 & 1.45$\pm$0.021 & 4.84$\pm$0.04$^5$ \\
M1-40 & 008.3 $-$01.1 & 1.38$\pm$0.013 & 0.09$\pm$0.001 & 1.9$\pm$0.09 & 0.35$\pm$0.005 & 0.82$\pm$0.011 & 5.15$\pm$0.04$^2$ \\
Th3-4 & 354.5 $+$03.3 & 1.56$\pm$0.015 & 0.42$\pm$0.004 & 1.2$\pm$0.06 & 0.63$\pm$0.008 & 1.45$\pm$0.021 & 4.68$\pm$0.04$^3$ \\
M3-38 & 356.9 $+$04.4 & 1.92$\pm$0.029 & 0.06$\pm$0.001 & 1.1$\pm$0.05 & 0.45$\pm$0.006 & 1$\pm$0.014 & 5.09$\pm$0.04$^3$ \\
H1-43 & 357.1 $-$04.7 & 2.22$\pm$0.022 & 0.09$\pm$0.001 & 1.1$\pm$0.05 & $-$1.61$\pm$0.022 & -- \,  \,  --& 4.26$\pm$0.04$^3$\\
H1-40 & 359.7 $-$02.6 & 2.07$\pm$0.020 & 0.27$\pm$0.002 & 1.2$\pm$0.06 & 0.12$\pm$0.001 & 0.84$\pm$0.011 & 4.85$\pm$0.04$^6$ \\
\hline
   \end{tabular}
\end{table*}

In Fig. \ref{12Temp}, we also present plots of the H$_2$ line flux 
against and the integrated intensity of the PAH feature, 
for both observed (panel c) and rescaled to bulge distance (panel d) values, in these panels the error bars are smaller than the symbols. 
 Collectively, these plots reinforce the notion that GBPNe have stronger 
H$_2$ (12.4$\upmu$m) emission than disk PNe, just as determined above by relating the H$_2$ (12.4$\upmu$m) flux to the angular size of the PNe.

In Fig. \ref{12Temp}, we plot the integrated flux of the 11.3$\upmu$m PAH band 
(scaled to the distance of the bulge, panel e) and the H$_2$ (12.4$\upmu$m) integrated
 flux (again scaled to the distance of the bulge, panel f) versus the effective 
temperature of the central star, also for these panels the error bars are smaller than the symbols. Neither line shows a strong correlation with the 
effective temperature, but the a slight anti-correlation with H$_2$ (12.4$\upmu$m) 
line flux suggests that H$_2$ is being photodissociated by the central star. It is
 also important to note that we do not observe any correlation between the the 
strengths of the H$_2$ line and the PAH band, implying that they come from different
 regions of the nebula. As such, in order to account for the relationship between 
the central star temperature and strength of the H$_2$ (12.4$\upmu$m) feature, 
we infer that the H$_2$ emitting region lies closer to the central star than that corresponding to the PAHs.

\section{Discussion}

Our observations confirm the presence of PAHs in the tori of 9 PNe in our sample, in the case of Th3-4 and H1-40 they were not resolved and M3-15 was the only object that showed no emission in either PAH1 nor PAH2 filters.
 This strongly supports the hypothesis that the chemical pathway to PAH formation occurs within high density regions.
 These tori remain molecular for some time after the ionisation of the nebula has started, due to the trapping of the
 ionisation front and due to  shielding effects of the dust column density (\citealp{woods05}).
 A mini-photon-dominated region (PDR) is expected in these molecular regions (\citealp{phillips10}), 
as they are slowly overrun by the advancing ionisation front.
A mini-PDR within a dense torus is a highly plausible location for the formation PAHs. By combining 
narrow band images of  PAH and [SIV] emission lines, we demonstrate that the PAHs  are generally located
 in the torus of each PNe (H1-43 is the only PNe that displays emission from PAHs in the outflows as well 
as in the torus), consistent with this hypothesis.

This mixed chemistry objects which comprise our sample are distinct from other dual dust
 chemistry objects such as the Red Rectangle \citep{waters98} and PNe with WC central stars \citep{waters}.
 These objects are C-rich and display oxygen features from earlier mass-loss, when the star was still O-rich.
 The O-rich material in these objects is thought to be present in a large disc, while the PAHs should be
 present in the outflows. \cite{perea09} proposed that the mixed chemistry in GBPNe was due to a recent
 thermal pulse, which made the star turn from O-rich to C-rich. If this were the case, one would expect
 to see carbon-rich material in the outflows of the nebula. This is only seen in H1-43 (and even then it is barely detected), 
while for all the other sources, the PAHs are concentrated in the torus. 

\cite{me11} present chemical models which predict that long C-chains would form on the surface of dense structures, when CO
 is photodissociated by a strong radiation field at an A$_V$$\sim$4--6. The observations presented here indicate that PAHs
 are very likely produced at the outer edges of the dusty structures tori in the objects that comprise our sample. These
 regions should be dense enough for the required chemistry to occur, but also rarefied enough for the ionising radiation
 to penetrate. This scenario has been observed and modelled for several PNe, where clumpiness is taken into account, and
 it has been shown that some molecules can survive longer than others and the radiation field can penetrate to these dense
 structures (\citealp{redman03}).
The PNe presented in this work are from a distinct population of mixed chemistry objects, as they are still clearly O-rich.
 The observations we present thus confirm that C-rich material, which is the building blocks of life, can be formed even in O-rich environments.\\

We have also shown no clear correlation between the H$_2$ emission line and the
 11.3$\upmu$m PAH feature. This is indicative that the excited molecular material and the PAHs reside in different regions 
of the nebulae. H$_2$ has been observed in more than 70 PNe, with observations showing that, in general, these
 objects have in a high or moderate N/O abundance ratio and bipolar morphology. Those PNe with detections of
 H$_2$ include both young and evolved PNe, suggesting that the molecule can survive over significant fraction
 of the nebular lifetime \citep{kimura12}. However, we find a slight anti-correlation between the 12.4$\upmu$m
 H$_2$ line strength and the temperature of the central star, indicating H$_2$ is photodissociated by the central star.

It is intriguing to note that planets have recently been detected around new post-common-envelope systems,
 with great uncertainty surrounding their origin \citep{beuer10}.  The formation of dusty/dense tori in PNe
 has been strongly associated with the central star passing through a common-envelope phase 
\citep{miszalski09,jones10,corradi11}, and with these structures now being clearly shown to 
offer the possibility to form complex molecular material, it offers the prospect of forming
 new circumbinary planets very quickly after the common-envelope phase \citep[as opposed to
 requiring them to survive the common-envelope phase which necessitates extreme fine-tuning
 of gravitation and drag forces experience by the planet inside the envelope]{beuer10}.

\section{Conclusions}

The main results and conclusions of this paper can be summarised as follows:
\begin{itemize}
\item Emission of the PAH features at 8.6$\upmu$m and 11.3$\upmu$m and the [SIV] line were imaged using the VISIR instrument in the {\it VLT}.
We find dense, dusty, toroidal structures in 10 (out of 11) objects we observed. We detected PAHs in the torus of 8 of them (we could not resolved the torus in Th3-4 and H1-43), this finding confirms the proposed PAH formation scenario whereby CO is photodissociated as described by the models presented in \citet{me11}.
\item  For most of the objects the [SIV] line shows emission in an inner region than the PAHs. One would expect these emission lines to come from an inner region, where the elements can recombine, while PAHs would form just outside this region, where some UV-photons penetrate and dissociate CO.  The free C then will aggregate in these regions, leading to the formation of the PAHs.
\item We found that the 12.4$\upmu$m H$_2$ line anti-correlates with the temperature of the central star (i.e. PNe with hotter central stars, display lower levels of H$_2$ emission). This shows that the H$_2$ is being photodissociated by the UV photons of the central star.
\end{itemize}

\section*{Acknowledgments}
This work is based on observations made with the VISIR/{\it VLT} from ESO
program 087.D-0270 , PI Lagadec, E. Also observations made with the {\it Spitzer} Space Telescope, which is operated by the Jet Propulsion Laboratory, California Institute of
Technology under a contract with NASA. LGR and AZ acknowledge the support of CONACyT (Mexico). Astrophysics at Manchester is supported by  grants from the STFC. This work
was co-funded under the Marie Curie Actions of the European Commission
(FP7-COFUND). It is a pleasure to thank the anonymous referee for useful comments and suggestions that improved the paper.

\bsp

\label{lastpage}


\begin{thebibliography}{99}
\bibitem[\protect\citeauthoryear{Acker et al.}{2002}]{acker02} Acker A., Gesicki K., Grosdidier Y., Durand S., 2002, A\&A, 384, 620 
\bibitem[\protect\citeauthoryear{Acker \& Neiner}{2003}]{acker03} Acker A., Neiner C., 2003, A\&A, 403, 659 
\bibitem[\protect\citeauthoryear{Azzopardi, Lequeux, \& Rebeirot}{1988}]{azzo} Azzopardi M., Lequeux J., Rebeirot E., 1988, A\&A, 202, L27 
\bibitem[\protect\citeauthoryear{Beuermann et al.}{2010}]{beuer10} {{Beuermann} K., {Hessman} F.~V., {Dreizler} S., {Marsh} T.~R., {Parsons} S.~G., {Winget} D.~E., {Miller} G.~F., {Schreiber} M.~R., {Kley} W., {Dhillon} V.~S., {Littlefair} S.~P., {Copperwheat} C.~M., {Hermes}, J.~J.}, 2010, A\&A, 521, 60L 
\bibitem[\protect\citeauthoryear{Cohen et al.}{1999}]{cohen99} Cohen M., Barlow M.~J., Sylvester R.~J., Liu X.-W., Cox P., Lim T., Schmitt B., Speck A.~K., 1999, ApJ, 513, L135 
\bibitem[\protect\citeauthoryear{Cohen et al.}{2002}]{cohen02} Cohen M., Barlow M.~J., Liu X.-W., Jones A.~F., 2002, MNRAS, 332, 879
\bibitem[\protect\citeauthoryear{Corradi et al.}{2011}]{corradi11} Corradi R. L. M., Sabin L., Miszalski B., Rodr\'iguez-Gil P., Santander-Garc\'ia M., Jones D., Drew J. E., Mampaso A., Barlow M. J., Rubio-D\'iez M. M., Casares J., Viironen K., Frew D. J., Giammanco C., Greimel R., Sale S. E., 2011, MNRAS, 410, 1349
\bibitem[Garc{\'{\i}}a-Hern{\'a}ndez et al.(2006)]{garcia06} Garc{\'{\i}}a-Hern{\'a}ndez, D.~A., Manchado, A., Garc{\'{\i}}a-Lario, P., et al.\ 2006, \apj, 640, 829
\bibitem[\protect\citeauthoryear{Gesicki \& Zijlstra}{2000}]{gesicki00} Gesicki K., Zijlstra A.~A., 2000, A\&A, 358, 1058 
\bibitem[\protect\citeauthoryear{Gesicki et al.}{2006}]{gesicki06} Gesicki K., Zijlstra A.~A., Acker A., G{\'o}rny S.~K., Gozdziewski K., Walsh J.~R., 2006, A\&A, 451, 925 
\bibitem[\protect\citeauthoryear{Gesicki \& Zijlstra}{2007}]{gesicki07} Gesicki K., Zijlstra A.~A., 2007, A\&A, 467, L29 
\bibitem[\protect\citeauthoryear{Groves, Nefs, \& Brandl}{2008}]{groves08} Groves B., Nefs B., Brandl B., 2008, MNRAS, 391, L113 
\bibitem[\protect\citeauthoryear{Gutenkunst et al.}{2008}]{guten} Gutenkunst S., Bernard-Salas J., Pottasch S.~R., Sloan G.~C., Houck J.~R., 2008, ApJ, 680, 1206 
\bibitem[\protect\citeauthoryear{Guzman-Ramirez et al.}{2011}]{me11} Guzman-Ramirez L., Zijlstra A.~A., N{\'{\i}}chuim{\'{\i}}n R., Gesicki K., Lagadec E., Millar T.~J., Woods P.~M., 2011, MNRAS, 414, 1667 
\bibitem[\protect\citeauthoryear{Herwig}{2005}]{herwig} Herwig F., 2005, ARA\&A, 43, 435 
\bibitem[\protect\citeauthoryear{Jones et al.}{2010}]{jones10} Jones D., Lloyd M., Santander-Garc\'ia M., L\'opez, J. A., Meaburn J., Mitchell D. L., O'Brien T. J., Pollacco D., Rubio-D\'iez M. M., Vaytet N. M. H., 2010, MNRAS, 408, 2312 
\bibitem[\protect\citeauthoryear{Kimura, Gruenwald, \& Aleman}{2012}]{kimura12} Kimura R.~K., Gruenwald R., Aleman I., 2012, A\&A, 541, A112 
\bibitem[\protect\citeauthoryear{Kwok}{2000}]{kwok} Kwok S., 2000, oepn.book
\bibitem[\protect\citeauthoryear{Lagadec et al.}{2011}]{lagadec11} Lagadec E., et al., 2011, MNRAS, 417, 32 
\bibitem[\protect\citeauthoryear{Lagage et al.}{2004}]{lagage04} Lagage P.~O., et al., 2004, Msngr, 117, 12
\bibitem[Leger \& Puget(1984)]{leger84} Leger, A., \& Puget, J.~L.\ 1984, \aap, 137, L5 
 \bibitem[\protect\citeauthoryear{Miszalski et al.}{2009}]{miszalski09} Miszalski B., Acker A., Parker Q. A., Moffat A. F. J., 2009, A\&A, 505, 249
\bibitem[\protect\citeauthoryear{Perea-Calder{\'o}n et al.}{2009}]{perea09} Perea-Calder{\'o}n J.~V., Garc{\'{\i}}a-Hern{\'a}ndez D.~A., Garc{\'{\i}}a-Lario P., Szczerba R., Bobrowsky M., 2009, A\&A, 495, L5 
\bibitem[\protect\citeauthoryear{Phillips}{2003}]{phillips03} Phillips J.~P., 2003, MNRAS, 344, 501
\bibitem[\protect\citeauthoryear{Phillips}{2010}]{phillips10} Phillips J. P., Cuesta L. C., Ramos-Larios G., 2010, MNRAS, 1448 
\bibitem[\protect\citeauthoryear{Phillips \& M{\'a}rquez-Lugo}{2011}]{phillips11} Phillips J.~P., M{\'a}rquez-Lugo R.~A., 2011, RMxAA, 47, 83 
\bibitem[\protect\citeauthoryear{Redman et al.}{2003}]{redman03} Redman M.~P., Viti S., Cau P., Williams D.~A., 2003, MNRAS, 345, 1291 
\bibitem[\protect\citeauthoryear{Rees \& Zijlstra}{2013}]{rees13} Rees B., Zijlstra A.~A., 2013, MNRAS, 435, 975 
\bibitem[\protect\citeauthoryear{Sahai et al.}{2011}]{sahai11} Sahai, R., Morris, M.~R., \& Villar, G.~G.\ 2011, \aj, 141, 134 
\bibitem[\protect\citeauthoryear{Sylvester et al.}{1999}]{sylvester} Sylvester R.~J., Kemper F., Barlow M.~J., de Jong T., Waters L.~B.~F.~M., Tielens A.~G.~G.~M., Omont A., 1999, A\&A, 352, 587 
\bibitem[\protect\citeauthoryear{Tyndall et al.}{2012}]{tyndall12} Tyndall A.~A., Jones D., Lloyd M., O'Brien 
T.~J., Pollacco D., 2012, MNRAS, 422, 1804 
\bibitem[\protect\citeauthoryear{Uttenthaler et al.}{2012}]{stephan} Uttenthaler S., Blommaert J.~A.~D.~L., Lebzelter T., Ryde N., Wood P.~R., Schultheis M., Aringer B., 2012, EPJWC, 19, 6009 
\bibitem[\protect\citeauthoryear{Uttenthaler et al.}{2007}]{stephan2} Uttenthaler S., Hron J., Lebzelter T., Busso M., Schultheis M., K{\"a}ufl H.~U., 2007, A\&A, 463, 251 
\bibitem[\protect\citeauthoryear{Vassiliadis \& Wood}{1993}]{vassi} Vassiliadis E., Wood P.~R., 1993, ApJ, 413, 641 
\bibitem[\protect\citeauthoryear{Waters et al.}{1998}]{waters98} Waters L.~B.~F.~M., et al., 1998, A\&A, 331, L61 
\bibitem[\protect\citeauthoryear{Waters et al.}{1998}]{waters} Waters L.~B.~F.~M., et al., 1998, Natur, 391, 868 
\bibitem[\protect\citeauthoryear{Woods et al.}{2005}]{woods05} Woods P. M., Nyman L., Sch$\ddot{\rm o}$ier F. L., Zijlstra A. A., Millar T. J., Olofsson H., 2005, A\&A, 429, 977
\bibitem[\protect\citeauthoryear{Zijlstra et al.}{1991}]{zijlstra91} Zijlstra A.~A., Gaylard M.~J., te Lintel Hekkert P., Menzies J., Nyman L.-A., Schwarz H.~E., 1991, A\&A, 243, L9 
\end{thebibliography}
\end{document}